\def\spose#1{\hbox to 0pt{#1\hss}}
\def\simlt{\mathrel{\spose{\lower 3pt\hbox{$\mathchar"218$}}
     \raise 2.0pt\hbox{$\mathchar"13C$}}}
\def\simgt{\mathrel{\spose{\lower 3pt\hbox{$\mathchar"218$}}
     \raise 2.0pt\hbox{$\mathchar"13E$}}}
\def\simpropto{\mathrel{\spose{\lower 3pt\hbox{$\mathchar"218$}}
     \raise 2.0pt\hbox{$\propto$}}}

\documentclass[twocolumn,aps,prd,nofootinbib,showpacs]{revtex4-1}
\usepackage{amsmath,graphicx,bm,color}
\begin{document}
\title{The Epoch of Reionization Window: I. Mathematical Formalism}

\author{Adrian Liu}
\email{acliu@berkeley.edu}
\affiliation{Department of Astronomy, UC Berkeley, Berkeley, CA 94720, USA}
\affiliation{Berkeley Center for Cosmological Physics, UC Berkeley, Berkeley, CA 94720, USA}
\author{Aaron R. Parsons}
\affiliation{Department of Astronomy, UC Berkeley, Berkeley, CA 94720, USA}
\affiliation{Radio Astronomy Laboratory, UC Berkeley, Berkeley, CA 94720, USA}
\author{Cathryn M. Trott}
\affiliation{International Centre for Radio Astronomy Research, Curtin University, Bentley, WA, Australia}
\affiliation{ARC Centre of Excellence for All-Sky Astrophysics (CAASTRO), Curtin University, Bentley WA, Australia}

\date{\today}

\newcommand{\apjs}{Astrophys. J. Suppl. Ser.}
\newcommand{\aj}{Astron. J.}
\newcommand{\mnras}{Mon. Not. R. Astron. Soc.}
\newcommand{\apjl}{Astrophys. J. Lett.}
\newcommand{\aap}{Astron. Astrophys.}
\newcommand{\pasa}{PASA}
\newcommand{\physrep}{Phys. Rep.}
\newcommand{\araa}{Annu. Rev. Astron. Astrophys.}


\begin{abstract}
The $21\,\textrm{cm}$ line provides a powerful probe of astrophysics and cosmology at high redshifts, but unlocking the potential of this probe requires the robust mitigation of foreground contaminants that are typically several orders of magnitude brighter than the cosmological signal.  Recent simulations and observations have shown that the smooth spectral structure of foregrounds combines with instrument chromaticity to contaminate a ``wedge"-shaped region in cylindrical Fourier space.  While previous efforts have explored the suppression of foregrounds within this wedge, as well as the avoidance of this highly contaminated region, all such efforts have neglected a rigorous examination of the error statistics associated with the wedge.  Using a quadratic estimator formalism applied to the interferometric measurement equation, we provide a framework for such a rigorous analysis (incorporating a fully covariant treatment of errors).  Additionally, we find that there are strong error correlations at high spatial wavenumbers that have so far been neglected in sensitivity derivations.  These error correlations substantially degrade the sensitivity of arrays relying on contributions from long baselines, compared to what one would estimate assuming uncorrelated errors.
\end{abstract}

\maketitle
\section{Introduction}

Modern cosmological observations have produced exquisite constraints on both the initial and final conditions of structure formation in our Universe.  Initial conditions have now been probed to high significance with a large number of cosmic microwave background experiments \cite{HinshawEtAl2013,Planck}, while at low redshifts, a combination of galaxy surveys and traditional astronomical measurements provide the final conditions \cite{SDSS}.  Still missing from these direct observations, however, are the intermediate epochs that bridge the gap between early and late times.  For example, despite tremendous recent progress in high-redshift galaxy observations, details regarding the formation of the first luminous objects and their effects on the intergalactic medium (IGM) during the Epoch of Reionization (EoR) remain uncertain.

In the next few years, direct observations of the EoR will be made possible by measurements of the redshifted $21\,\textrm{cm}$ hyperfine transition of neutral hydrogen (see, e.g., Refs. \cite{Furlanetto2006,Morales2010,Pritchard2012,AviBook} for reviews).  At the relevant redshifts, the intensity field of the $21\,\textrm{cm}$ brightness temperature depends on a rich variety of different astrophysical effects, such as fluctuations in the ionization and spin states of the IGM, as well as cosmological quantities such as the underlying dark matter density field and peculiar velocity gradients.  A map of the $21\,\textrm{cm}$ intensity field at redshifts $z \sim 6$ and above would therefore be a rich probe of EoR physics, including the nature of the first luminous sources (such as their typical mass and luminosity scales), their ionizing and heating efficiency, and feedback processes on the IGM, among other effects.  Such a mapping can be accomplished in three dimensions, since the spectral nature of a $21\,\textrm{cm}$ measurement provides redshift (and therefore line-of-sight distance) information, while the angular directions are mapped using traditional imaging.  Because of this, the $21\,\textrm{cm}$ line allows access to a large fraction of our Universe's comoving volume, potentially allowing futuristic measurements to move beyond astrophysics and into the measurement of fundamental cosmological parameters \cite{McQuinn2006,Bowman2007,Mao2008}.

There are currently a number of experiments aimed at mapping the fluctuations of the cosmological $21\,\textrm{cm}$ signal, including the Giant Metrewave Radio Telescope Epoch of Reionization experiment (GMRT-EoR \cite{Paciga2013}), the Low Frequency Array (LOFAR \cite{Yatawatta2013}), the Murchison Widefield Array (MWA \cite{Tingay2013}), and the Donald C. Backer Precision Array for Probing the Epoch of Reionization (PAPER \cite{Parsons2010}).  These interferometer arrays have yet to make a positive detection of the cosmological signal, with the primary challenges being foreground contamination and the high sensitivity requirements.  To increase sensitivity, these experiments are primarily targeting binned, statistical measures of the brightness temperature field such as the power spectrum.  Recent progress has resulted in a number of increasingly stringent upper limits \cite{Paciga2013,Dillon2014,Parsons2013}, and proposed next-generation instruments such as the Hydrogen Epoch of Reionization Array (HERA \cite{Pober2014}) and the Square Kilometer Array (SKA \cite{Mellema2013}) promise to yield extremely high significance measurements.

In addition to achieving the required sensitivity, observations targeting the redshifted $21\,\textrm{cm}$ line must also contend with foreground contaminants. In the relevant frequency ranges (roughly $\sim 100$ to $200\,\textrm{MHz}$, corresponding to $z\sim13$ to $6$), there exist a large number of non-cosmological sources of radio emission that contaminate measurements.  These include sources such as the diffuse synchrotron radiation from our own Galaxy, as well as extragalactic point sources, whether they are bright and resolved or part of a dim and unresolved background.  The brightness temperature of foregrounds is expected to be $10^5$ times greater than theoretical expectations for the amplitude of the cosmological signal.  A detection of the reionization power spectrum will therefore be challenging without a robust foreground mitigation strategy.

Historically, cosmic microwave background (CMB) experiments have had to deal with similar problems of foreground contamination.  However, strategies for foreground cleaning that have been developed for the CMB cannot be directly applied to $21\,\textrm{cm}$ cosmology for two reasons.  First, CMB experiments typically operate at higher frequencies, where foregrounds are not as bright.  In fact, microwave-frequency foregrounds are subdominant to the CMB away from the Galactic plane.  In addition, CMB experiments measure anisotropies over a two-dimensional surface, with different observation frequencies providing consistency-checks and a set of redundant measurements that can be used for foreground isolation.  The three-dimensional mapping of the $21\,\textrm{cm}$ line, on the other hand, contains unique cosmological information at every frequency, which makes it more difficult to remove foregrounds in a way that does not result in the loss of cosmological signal \cite{Liu2013}.

With CMB techniques unlikely to succeed without modification, a number of alternate foreground mitigation strategies have been suggested for $21\,\textrm{cm}$ cosmology.  These include spectral polynomial fitting \cite{Wang2006,Liu2009a,Bowman2009,Liu2009b}, Wiener filtering \cite{Gleser2008}, principal component analyses \cite{Paciga2011,Liu2012,Masui2013,Paciga2013}, non-parametric subtractions \cite{Harker2009,Chapman2012,Chapman2013}, Fourier-mode or delay filtering \cite{Petrovic2011, Parsons2012b}, frequency stacking \cite{Cho2012}, Karhunen-Lo\`{e}ve eigenmode projection \cite{Shaw2014a,Shaw2014b}, and inverse covariance weighting \cite{Liu2011,Dillon2013,Dillon2014}.  The vast majority of these approaches rely on the fact that foreground sources are expected to be spectrally smooth, while the cosmological EoR signal is expected to fluctuate rapidly with frequency \cite{Oh2003}.  The cosmological signal can therefore be extracted by isolating spectrally smooth components from the data.

\begin{figure}[t] 
	\centering 
	\includegraphics[width=.45\textwidth]{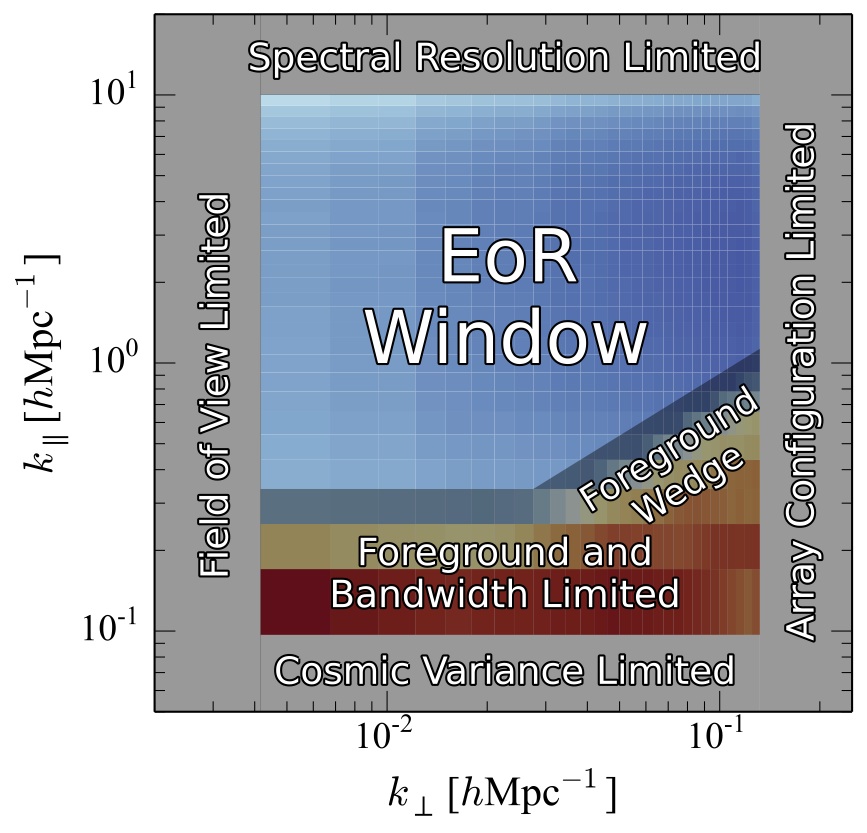}
	\caption{A schematic of the EoR window in the cylindrical $k_\perp k_\parallel$ Fourier plane.  At the lowest $k_\perp$, errors increase because of limits on an instrument's field-of-view.  High $k_\perp$ modes are probed by the longest baselines of an interferometer array, and the sensitivity drops to zero beyond $k_\perp$ scales corresponding to these baselines.  Spectral resolution limits the sensitivity at large $k_\parallel$.  The lowest $k_\parallel$ are in principle limited by cosmic variance, but in practice the larger concern is limited bandwidth and the foreground contamination, which intrinsically resides at low $k_\parallel$.  As one moves towards higher $k_\perp$, however, the foregrounds leak out to higher $k_\parallel$ in a characteristic shape known as the ``foreground wedge".  The remaining parts of the Fourier plane are thermal-noise dominated, allowing (with a large collecting area or a long integration time) a clean measurement of the power spectrum in this ``EoR window".}
	\label{fig:SchematicWedge}
\end{figure} 

Recently, however, a complication to this simple picture was realized, in what has been colloquially termed the ``foreground wedge".  Consider a cylindrically-binned power spectrum measurement, i.e. one where Fourier amplitudes are squared and binned in annuli specified by wavenumbers perpendicular to the line-of-sight $k_\perp$ and wavenumbers parallel to the line-of-sight $k_\parallel$.  Because the line-of-sight direction is equivalent to the spectral axis of an interferometer, one might have naively expected smooth spectrum foregrounds to be sequestered to only the lowest $k_\parallel$.  However, this neglects the fact that interferometers are inherently chromatic instruments, with a given baseline probing finer spatial scales (higher $k_\perp$) at higher frequencies.  This coupling of spectral and spatial information is sometimes coined mode-mixing, and results in  the leakage of information from low to high $k_\parallel$.  This effect is particularly pronounced at high $k_\perp$, where the modes are typically probed by longer baselines, which are more chromatic.  Putting everything together, theoretical studies and simulations \cite{Datta2010,Vedantham2012,Morales2012,Parsons2012b,Trott2012,Thyagarajan2013,Hazelton2013} have shown that foregrounds are expected to leak out of the lowest $k_\parallel$ into a characteristic ``wedge" that is schematically shown in Figure \ref{fig:SchematicWedge}.  Observations with PAPER and MWA have confirmed this basic picture \cite{Pober2013}, including its evolution with frequency \cite{Dillon2014}.

The foreground wedge is both a blessing and a curse.  At the sensitivity levels that have been achieved by current experiments, observations have seen a sharp drop-off in foregrounds beyond the wedge \cite{Pober2013}.  Theoretical calculations and simulations have shown that such a drop-off is the natural consequence of geometric limitations \cite{Parsons2012b}, provided the foregrounds are spectrally smooth.  If further integration reveals low-level foregrounds that are spectrally unsmooth, their influence will leak beyond what is typically labeled as the edge of the wedge.  However, if foregrounds continue to be reasonably smooth, the fact that physical considerations limit the extent of wedge implies that there must exist an ``EoR window": a region in Fourier space that is \emph{a priori} expected to be foreground-free.  The existence of the EoR window thus enables a relatively robust foreground avoidance strategy, where a detection of the power spectrum can be made simply by avoiding measurements within the wedge.  On the other hand, such a conservative approach forces one to work at higher $k$ than if the chromatic effects had not caused the wedge in the first place.  This is unfortunate because the ratio of the cosmological signal to instrumental noise typically peaks at low $k$, which means that if it were possible to work within the wedge (or to at least push back its boundaries a little), one would be able to make higher significance detections of the power spectrum.  Indeed, in Ref. \cite{Pober2014} it was suggested that working within the wedge can increase the detection significance anywhere from a factor of two to six (depending on the interferometer's configuration), with corresponding decreases in the error bars on astrophysical parameters.

Given the potentially high payoff associated with pushing back the influence of the wedge (or equivalently, enlarging the EoR window), it is important to have a statistically rigorous framework for describing the wedge.  In this paper, we provide just such a mathematical framework.  Since there already exists an extensive literature on the foreground wedge and the EoR window, it is worth summarizing the ways in which this paper builds upon and extends previous results.  Works such as Refs. \cite{Datta2010,Vedantham2012,Morales2012} describe instrumental simulations that take one particular realization of foregrounds and propagate them through a power spectrum estimation pipeline.  They therefore only probe the \emph{mean} power spectrum, and not the scatter (i.e. the errors) about this mean.  In Ref. \cite{Thyagarajan2013}, a statistical treatment of point source populations was considered.  While error bars were computed, off-diagonal error correlations (i.e. covariances) in the final measurements were neglected.  Potential error correlations are important particularly because Ref. \cite{Thyagarajan2013} considered the application of various tapering functions to their Fourier transforms, and certain choices can result in significant correlations between Fourier bins.  We consider full foreground covariances, a full treatment of instrumental effects (such as having a non-tophat beam), a full treatment of data analysis choices such as tapering functions, and a fully covariant propagation of errors.  We build upon Ref. \cite{Trott2012}, which made use of Monte Carlo methods to propagate errors.  Our treatment is more analytic, allowing us to capture the large dynamic range needed to accurately compute the error statistics in a measurement where the foregrounds are many orders of magnitude brighter than the signal.  This is made computationally feasible by our use of the delay spectrum approach introduced in Ref. \cite{Parsons2012b}, where input frequency spectra are sorted into a set of time-delay $\tau$ modes via a per-baseline Fourier transform.  However, unlike in some works where the delays are used as an \emph{approximation} for line-of-sight Fourier modes (an assumption that is only valid for short baselines, as we will discuss in Section \ref{sec:InstrumentResponse}), we use delays strictly as a convenient choice of \emph{basis}.  This basis makes it computationally possible for us to deal directly with visibilities in our formalism (bypassing any mapmaking steps), which avoids gridding artifacts in our numerical results.  We also take into full account the correlations between partially overlapping baselines, and therefore rigorously treat the possible complications that were highlighted in Ref. \cite{Hazelton2013}.  A related treatment pertaining to lower-redshift $21\,\textrm{cm}$ intensity mapping experiments (though focusing less on the details of the wedge) can be found in Ref. \cite{Shaw2014b}.  While our fiducial calculations are centered around instruments targeting the EoR, the techniques developed in this paper are equally applicable to cosmological $21\,\textrm{cm}$ at lower (or higher) redshifts.

We accomplish our goals by making use of the quadratic estimator formalism, which was adapted for $21\,\textrm{cm}$ power spectrum measurements in Refs. \cite{Liu2011,Dillon2013}, and applied to real data in Ref. \cite{Dillon2014}.  However, appropriate ``wedge effects" were not incorporated into the formalism, an omission that we rectify in this paper.  Placing everything in the quadratic estimator formalism enables a systematic computation of the aforementioned error statistics, as well as a systematic study of the optimality (or lack thereof) of various power spectrum estimators.  In a sequel paper (Ref. \cite{Liu2014b}, henceforth ``Paper II"), we will take advantage of this to examine the extent to which statistical methods can enlarge the EoR window.

With our fully covariant treatment, we find that the wedge is not simply a region of large foreground errors and biases, but also as a marker for error correlations: at $k_\perp$ values where the wedge is a dominant effect, the errors tend to be strongly correlated.  With strongly correlated errors, the number of independently measurable Fourier modes is reduced, suggesting that previous sensitivity estimates (such as those in Refs. \cite{Parsons2012a,Beardsley2013,Pober2014}) may be overly optimistic, particularly for arrays that make use of long baselines (such as LOFAR or GMRT).  In fact, the rewards for working within the wedge may be overrated as a result of this, but of course this cannot be quantified without a rigorous way to compute the error statistics of the wedge---hence the present paper.

The rest of this work is organized as follows.  In Section \ref{sec:InstrumentResponse} we examine the measurement equation of an interferometer in detail, paying special attention to chromatic effects.  This provides a first non-covariant preview of the foreground wedge, which we generalize to an approximate, but fully covariant description in Section \ref{sec:basicEstAnalytic}, following a review of the quadratic estimator formalism in Section \ref{sec:QuadEst}.  In Section \ref{sec:basicEstNumerical} we discard the approximations made in Section \ref{sec:basicEstAnalytic} in a full numerical implementation of our formalism.  We summarize our conclusions in Section \ref{sec:Conclusions}.  Because a large number of mathematical quantities are defined in this paper, we provide dictionaries in Tables \ref{tab:Definitions} and \ref{tab:vectMatrixDefinitions} for the reader's convenience.

\begin{table*}
\caption{\label{tab:Definitions}Dictionary of scalars and functions.  The ``context" column gives equation references, typically either their defining equation or their first appearance in the text.}
\begin{ruledtabular}
\begin{tabular}{lll}
Quantity & Meaning/Definition &  Context \\
\hline
\multicolumn{3}{l}{Basic quantities}\\
$\mathbf{u}$ & Fourier dual to angular direction $\boldsymbol \theta$ & Eq. \eqref{eq:FourierDef} \\
$\eta$ & Fourier dual to $\nu$ (i.e. spectral wavenumber) & Eq. \eqref{eq:FourierDef} \\
$\tau$ & Delay, i.e. Fourier dual to $\nu$ (or spectral wavenumber) for a single baseline & Eq. \eqref{eq:DelayDef} \\
$V(\mathbf{b}, \nu)$ & Visibility measured by baseline $\mathbf{b}$ at frequency $\nu$  & Eq. \eqref{eq:measEqn} \\
$\widetilde{V} (\mathbf{b}, \tau)$ & Delay-space visibility by baseline $\mathbf{b}$ at delay $\tau$  & Eq. \eqref{eq:DelayDef} \\
\hline
\multicolumn{3}{l}{Instrumental parameters}\\
$\mathbf{b}$ & Baseline vector &  Eq. \eqref{eq:basicVis} \\
$\boldsymbol \theta_0 $ & Characteristic width of primary beam &  Eq. \eqref{eq:basicVis} \\
$A$ & Primary beam function &  Eq. \eqref{eq:basicVis} \\
$\widetilde{A}$ & Spatial Fourier transform of primary beam function &  Eq. \eqref{eq:FirstAtilde} \\
$\widetilde{A}_{b\parallel} $ & Profile of $\widetilde{A}$ parallel to baseline vector direction &  Eq. \eqref{eq:SepPrimaryBeam} \\
$\widetilde{A}_{b\perp} $ & Profile of $\widetilde{A}$ perpendicular to baseline vector direction &  Eq. \eqref{eq:SepPrimaryBeam} \\
$B_\textrm{chan} $ & Frequency channel width &  Eq. \eqref{eq:measEqn} \\
$\gamma $ & Frequency channel profile &  Eq. \eqref{eq:measEqn} \\
$\widetilde{\gamma} $ & Fourier transform of frequency channel profile $\gamma$ &  Eq. \eqref{eq:vIntegratedGenVtilde} \\
$B_\textrm{band} $ & Bandwidth corresponding to depth of cosmological volume &  Eq. \eqref{eq:DelayDef} \\
$\Omega_\textrm{pp}$ & Integrated beam squared area & Eq. \eqref{eq:Omegapp} \\
$T_\textrm{sys}$ & System temperature & Eq. \eqref{eq:NoiseCovarSingelBl} \\
$t$ & Total integration time & Eq. \eqref{eq:NoiseCovarSingelBl} \\
$n(b)$ & Number baselines  & Eq. \eqref{eq:NoiseCovarSingelBl} \\
\hline
Sky: && \\
$I(\boldsymbol \theta, \nu)$ & Sky brightness temperature at angle $\boldsymbol \theta$ and frequency $\nu$ & Eq. \eqref{eq:basicVis} \\
$\widetilde{I}(\mathbf{u}, \eta)$ & Fourier transform of the sky temperature at angular wavenumber $\mathbf{u}$ and line-of-sight wavenumber $\eta$ & Eq. \eqref{eq:FourierDef} \\
 $P(u,\eta)$ & Cylindrically-binned power spectrum at angular wavenumber $u$ and line-of-sight wavenumber $\eta$ & Eq. \eqref{eq:PowerSpectrumAppendix} \\
 $C_\ell^X$ & Angular power spectrum of foregrounds,  & Eq. \eqref{eq:PowerSpectrumForegrounds} \\
 $\nu_c^X$ & Frequency coherence length of foregrounds & Eq. \eqref{eq:PowerSpectrumForegrounds} \\
 & (For both $C_\ell^X$ and $\nu_c^X$, $X=\textrm{diff}$ for diffuse Galactic emission and $X=\textrm{ps}$ for point sources) & \\
\hline
Data analysis: &  &\\
$\phi$ & Bandpass or tapering function & Eq. \eqref{eq:DelayDef} \\
$\widetilde{\phi}$ & Fourier transform of bandpass/tapering function $\phi$ & Eq. \eqref{eq:shortBlVtilde} \\
$h(\mathbf{u}, \eta; \mathbf{b}, \tau)$ & Visibility response at delay $\tau$ of baseline $\mathbf{b}$ to sky mode on spatial scale $\mathbf{u}$ and spectral scale $\eta$.& Eq. \eqref{eq:hDef} \\
$g(u, \eta; b, \tau)$ & Same as $h(\mathbf{u}, \eta; \mathbf{b}, \tau)$ but integrated over direction on $uv$ plane perpendicular to baseline vector & Eq. \eqref{eq:GeeDef}
\end{tabular}
\end{ruledtabular}
\end{table*}

\begin{table*}
\caption{\label{tab:vectMatrixDefinitions}Dictionary of vectors and matrices.  The quantities shown here are grouped into three categories: those that exist in the vector space of the visibility measurements (indexed, for example, by baseline and delay), those that exist in the vector space of bandpowers (indexed by Fourier wavenumbers), and those that bridge the two spaces.  In the column giving the length/dimensions, $N_\textrm{bl}$ denotes the number of baselines, $N_\nu$ the number of frequency channels, and $N_\textrm{bands}$ the number of bins in Fourier space (i.e. the number of bandpowers).}
\begin{ruledtabular}
\begin{tabular}{lllll}
Quantity & Components & Meaning/Definition & Length/Dimensions  & Context \\
\hline
\multicolumn{5}{l}{Quantities in measurement space}\\
$\mathbf{x}$& $\mathbf{x}_i$ & Serialized data vector of visibilities & $N_\textrm{bl}N_{\nu}$ & Eq. \eqref{eq:xvectdef} \\
$\mathbf{C}$ & $\mathbf{C}_{ij}$ & Total covariance matrix $\mathbf{C} \equiv \langle \mathbf{x} \mathbf{x}^\dagger \rangle$ & $N_\textrm{bl}N_{\nu} \times N_\textrm{bl}N_{\nu}$ & Eq. \eqref{eq:C=N+S} \\
$\mathbf{N}$ & $\mathbf{N}_{ij}$ & Noise covariance matrix  & $N_\textrm{bl}N_{\nu} \times N_\textrm{bl}N_{\nu}$ & Eq. \eqref{eq:C=N+S} \\
$\mathbf{S}$ & $\mathbf{S}_{ij}$ & Signal covariance matrix  & $N_\textrm{bl}N_{\nu} \times N_\textrm{bl}N_{\nu}$ & Eq. \eqref{eq:genCovar} \\
\hline
\multicolumn{5}{l}{Quantities in power spectrum space}\\
 $\mathbf{p}$ &  $p_\alpha$ & Serialized vector of true bandpowers (i.e. power & $N_\textrm{bands}$ & Eq. \eqref{eq:CovarDecomp} \\
   &  &  spectrum at various grid points in Fourier space) &  & \\
$\widehat{\mathbf{p}}$ &    $\widehat{p}_\alpha$ & Estimator for bandpowers derived from measurements & $N_\textrm{bands}$ & Eq. \eqref{eq:GenQuadEst} \\
$\boldsymbol \Sigma$ & $\boldsymbol \Sigma_{\alpha \beta}$ &  Error covariance of estimated bandpowers,  & $N_\textrm{bands} \times N_\textrm{bands}$ & Eq. \eqref{eq:matrixErrorCovar} \\
& & i.e., $\boldsymbol \Sigma_{\alpha \beta} \equiv \langle \widehat{p}_\alpha \widehat{p}_\beta \rangle - \langle \widehat{p}_\alpha \rangle \langle \widehat{p}_\beta \rangle $ & & \\
$\overline{\boldsymbol \Sigma}$ & $\overline{\boldsymbol \Sigma}_{\alpha \beta}$ & Error correlation of estimated bandpowers & $N_\textrm{bands} \times N_\textrm{bands}$ & Eq. \eqref{eq:ErrorCorr} \\
$\mathbf{W}$ & $\mathbf{W}_{\alpha \beta}$ & Window function matrix & $N_\textrm{bands} \times N_\textrm{bands}$ & Eq. \eqref{eq:EstWindAndBias} \\
$\mathbf{b}$ & $\mathbf{b}_\alpha$ & Power spectrum estimator bias & $N_\textrm{bands}$ & Eq. \eqref{eq:bias} \\
$\mathbf{M}$ & $M_\alpha$ & Power spectrum estimator normalization & $N_\textrm{bands}$ & Eq. \eqref{eq:basicEstEalpha} \\
\hline
\multicolumn{5}{l}{Hybrid quantities}\\
 $\mathbf{E}_\alpha$ & $(\mathbf{E}_\alpha)_{ij}$ & Estimator matrix for quadratic estimator of & $N_\textrm{bl}N_{\nu} \times N_\textrm{bl}N_{\nu}$ & Eq. \eqref{eq:GenQuadEst} \\
 &  & of bandpower $p_\alpha$, i.e., $\widehat{p}_\alpha = \mathbf{x} \mathbf{E}_\alpha \mathbf{x}$ & for each $\alpha = 1$ to $N_\textrm{bands}$ & \\
  $\mathbf{C}_{,\alpha}$ &   $(\mathbf{C}_{,\alpha})_{ij}$ & Response of total covariance to the $\alpha$th bandpower, & $N_\textrm{bl}N_{\nu} \times N_\textrm{bl}N_{\nu}$ & Eq. \eqref{eq:CcommaAlpha} \\
   &  &  i.e., $\mathbf{C}_{,\alpha} \equiv \partial \mathbf{C} / \partial p_\alpha$ & for each $\alpha = 1$ to $N_\textrm{bands}$ & \\
\end{tabular}
\end{ruledtabular}
\end{table*}

\section{A non-covariant preview of the foreground wedge}
\label{sec:InstrumentResponse}
In this section, we will derive the Fourier-space foreground wedge from first principles.  We begin with the visibility $V$ measured by a baseline $\mathbf{b}$ at frequency $\nu$:
\begin{equation}
\label{eq:basicVis}
V(\mathbf{b}, \nu) = \int  I (\boldsymbol \theta, \nu)  A \left( \frac{\boldsymbol \theta}{\theta_0} \right)\exp \left( - i 2 \pi \frac{\nu}{c} \mathbf{b} \cdot \boldsymbol \theta \right)  d^2 \mathbf{\theta},
\end{equation}
where $\boldsymbol \theta$ is the angular sky position, $I (\boldsymbol \theta , \nu)$ is the sky temperature and $A(\boldsymbol \theta / \theta_0 )$ is the primary beam,\footnote{In general, the primary beam will depend on frequency, although for some instruments (such as PAPER) the antennas are intentionally designed to minimize the frequency-dependence of the beam \cite{Parsons2010}.  In this paper, we will neglect the frequency-dependence, because our goal is not to provide results pertaining to a particular instrument, but instead to provide a rigorous understanding of how foregrounds enter an interferometric power spectrum measurement.  Including a frequency-dependent beam makes many of our analytic manipulations more difficult, which obscures the key physical effects that give rise to the foreground wedge.  We note, however, that our general strategy of incorporating an interferometer's measurement equation into the quadratic estimator formalism is one that is capable of including frequency-dependent beams, albeit at a slightly greater computational cost.} with $\theta_0$ denoting its characteristic width.  For notational simplicity in this section, we will omit the instrumental noise contribution to the visibility, but of course it is always implicitly present.  Our (arbitrary) convention for $A$ is that it is dimensionless and is normalized so that $A(0)=1$.  In what follows, we will see that the foreground wedge arises from the fact that the product of $\nu$ and $\boldsymbol \theta$ appears in the complex exponential.  Fourier transforms in $\boldsymbol \theta$ are therefore coupled to $\nu$ and vice versa, leading to the ``mode-mixing" phenomena coined in Ref. \cite{Morales2012} and ultimately the wedge.

To mimic the discreteness of frequency channels in a real instrument, we introduce a function $\gamma$ that describes the response of a single frequency channel.  Our measurement equation therefore becomes
\begin{eqnarray}
\label{eq:measEqn}
V(\mathbf{b}, \nu) = \int   \frac{d\nu^\prime}{B_\textrm{chan}} && d^2 \boldsymbol{\theta} \, I (\boldsymbol \theta, \nu^\prime)  A \left( \frac{\boldsymbol \theta}{\theta_0} \right) \nonumber \\
&& \times \, \gamma\left( \frac{\nu-\nu^\prime}{B_\textrm{chan}} \right)   e^{ - i 2 \pi \frac{\nu^\prime}{c} \mathbf{b} \cdot \boldsymbol \theta }  ,
\end{eqnarray}
with $B_\textrm{chan}$ denoting the width of a frequency channel, and $\gamma$ normalized such that $\int_{-\infty}^\infty  \gamma (x) dx = 1$.

\begin{figure}[t!] 
	\centering 
	\includegraphics[width=.4\textwidth]{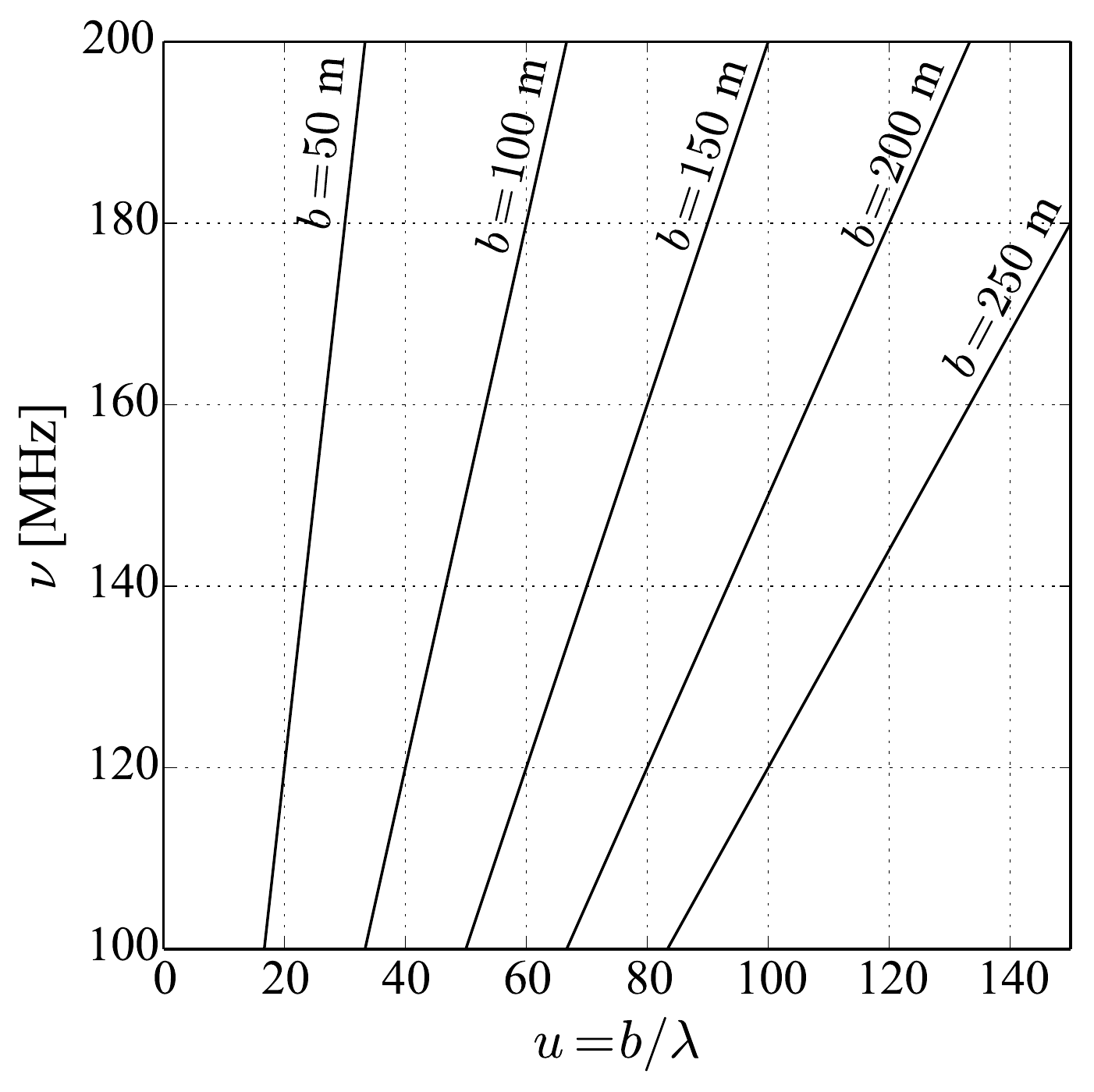}
	\caption{Angular Fourier coordinate $u$ probed by a variety of baseline lengths, plotted as a function of observing frequency.  Taking a Fourier transform of the frequency spectrum of data from a single baseline essentially amounts to taking a Fourier transform along a solid line in this figure.  This is a good approximation to taking a Fourier transform along the ``true" frequency axis for short baselines.}
	\label{fig:ChromaticBaselines}
\end{figure} 

So far, we have no choice in the matter, in that the results of  Eq. \eqref{eq:measEqn} are handed to us by the instrument.  Moving onto data analysis, however, there is considerable freedom as to how one proceeds.  For example, we may choose to move into delay-space, which is accomplished by taking the Fourier transform in frequency of the spectrum measured by a single baseline (a ``delay transform"):
\begin{equation}
\label{eq:DelayDef}
\widetilde{V} (\mathbf{b}, \tau) = \int V(\mathbf{b}, \nu) \phi \left( \frac{\nu - \nu_0}{B_\textrm{band}} \right) e^{-i 2 \pi \nu \tau} d\nu,
\end{equation}
where $\tau$ is the delay (with units of time), $B_\textrm{band}$ is the bandwidth over which we wish to compute a power spectrum, and $\nu_0$ is the central frequency of our band.  The function $\phi$ is normalized so that $\phi(0)=1$, and captures both the bandpass of our instrument and any tapering that one may wish to impose near the band edges.  (In what follows, we will therefore use the terms ``tapering function" and ``bandpass" interchangeably to describe $\phi$).  Precisely what form the edge tapering takes is a data analysis choice, and as shown in Ref. \cite{Thyagarajan2013}, choosing a good tapering function minimizes leakage of foregrounds in the Fourier plane.  Later on, we will extend the work of Ref. \cite{Thyagarajan2013} to self-consistently incorporate the effect that a tapering function has on error covariances.  Aside from the peculiarities of certain tapering functions, working in delay space is simply a change of basis, and as emphasized in Ref. \cite{Parsons2012b}, represents no loss of generality.  In later sections, we will find that delay space is a particular efficient basis to work in, one that makes many of the large matrices required for power spectrum estimation sparse (see Appendix \ref{appendix:Computational} for details).

At this point our data are in a basis specified by baseline vector $\mathbf{b}$ and delay mode $\tau$.  This basis closely approximates the Fourier basis that the power spectrum inhabits, but the correspondence is not exact and in general must be accounted for.  Concretely, suppose we let $\mathbf{u}$ be the Fourier dual of $\boldsymbol \theta$ and $\eta$ be the Fourier dual of $\nu$.  Since the transverse co-moving distance can be used to convert angles on the sky to transverse comoving distance, we have $\mathbf{u} \propto \mathbf{k}_\perp$ (the exact conversion is given in Appendix \ref{appendix:Fourier}, where we define our Fourier conventions).  Similarly, the observed frequency of a spectral line can be mapped to a co-moving line-of-sight distance, so $\eta \propto k_\parallel$.  At a particular frequency, a single baseline $\mathbf{b}$ of an interferometer roughly measures the spatial Fourier mode specified by wavenumber $\mathbf{u} = \mathbf{b} / \lambda$ [this can be seen by applying Parseval's theorem to Eq. \eqref{eq:basicVis} and assuming that the primary beam is wide].  Across different frequencies, however, we see that a single baseline probes different spatial Fourier modes.  A Fourier transform of a single baseline's spectrum [Eq. \eqref{eq:DelayDef}] is thus a Fourier transform along one of the solid lines in Figure \ref{fig:ChromaticBaselines}, where we show the chromaticity of various baselines.  As pointed out in Ref. \cite{Parsons2012b}, with short baselines it is an excellent approximation to say that $\eta \sim \tau$.  In this paper, we do not make this ``delay approximation", as we consider baselines of all lengths.\footnote{Note that the distinction between $\tau$ modes and $\eta$ modes implies that our tapering functions differ slightly from those examined in Ref. \cite{Thyagarajan2013}.  In this paper, the tapering functions are applied to the per-baseline Fourier transform, rather than the Fourier transform along the ``true" frequency axis of Figure \ref{fig:ChromaticBaselines}.}

To properly account for the mapping between the ``true" Fourier coordinates $(\mathbf{u}, \eta)$ to our visibilities expressed in the $(\mathbf{b}, \tau)$ basis, we can express the sky temperature $I (\boldsymbol \theta , \nu)$ in terms of its ``true" Fourier transform, $\widetilde{I} (\mathbf{u}, \eta)$:
\begin{equation}
\label{eq:FourierDef}
I (\boldsymbol \theta, \nu) = \int \widetilde{I} (\mathbf{u}, \eta) e^{i 2 \pi ( \mathbf{u} \cdot \boldsymbol \theta + \eta \nu)} d\eta \,d^2 u.
\end{equation}
Inserting  Eqs. \eqref{eq:measEqn} and \eqref{eq:FourierDef} into Eq. \eqref{eq:DelayDef} gives
\begin{widetext}
\begin{subequations}
\begin{eqnarray}
\widetilde{V} (\mathbf{b}, \tau) &=&\int d^2\mathbf{u} \, d\eta \widetilde{I} (\mathbf{u}, \eta) \int d^2 \boldsymbol \theta \,A \left( \frac{\boldsymbol \theta}{\theta_0} \right) \int d\nu \frac{d\nu^\prime}{B_\textrm{chan}} e^{-i 2 \pi \frac{\nu^\prime}{c} \mathbf{b} \cdot \boldsymbol \theta} \gamma\left( \frac{\nu-\nu^\prime}{B_\textrm{chan}} \right) \phi \left( \frac{\nu - \nu_0}{B_\textrm{band}} \right) e^{-i 2\pi \nu \tau} e^{i 2 \pi ( \mathbf{u} \cdot \boldsymbol \theta + \eta \nu^\prime)} \\
&=& \int d^2\mathbf{u} \, d\eta \widetilde{I} (\mathbf{u}, \eta) \theta_0^2  \int d\nu \frac{d\nu^\prime}{B_\textrm{chan}} \widetilde{A} \left[ \theta_0 \left( \mathbf{u} - \frac{\nu^\prime}{c} \mathbf{b} \right) \right] \gamma\left( \frac{\nu-\nu^\prime}{B_\textrm{chan}} \right) \phi \left( \frac{\nu - \nu_0}{B_\textrm{band}} \right) e^{i 2\pi(\nu^\prime \eta- \nu \tau)}  \label{eq:FirstAtilde}
\end{eqnarray}
\end{subequations}
The integral over $\nu^\prime$ can be simplified because $\widetilde{A}$ is a much broader function (with a characteristic width of $c\mathbf{b} / \theta_0$) than $\gamma$ (which has a characteristic width of $B_\textrm{chan}$).  For a typical EoR  experiment, one might have $\nu_0=150\,\textrm{MHz}$ and $B_\textrm{chan} = 50 \,\textrm{kHz}$.  Even with a rather conservative $\theta_0 = 1\,\textrm{radian}$, $\widetilde{A}$ is wider than $\gamma$ for any baseline shorter than $\sim 3000\lambda$.  With compact arrays giving the highest sensitivity \cite{Parsons2012a,Pober2014}, very few interferometers that are optimized for an EoR measurement have any relevant sensitivity on baselines this long.  As a result, $\widetilde{A}$ can be factored out of the integral and evaluated at $\nu^\prime = \nu$, leaving a Fourier transform over the $\nu^\prime$ variable.  Defining $\widetilde{\gamma}$ to be the Fourier transform of $\gamma$, this leaves
\begin{equation}
\label{eq:genVtilde}
\widetilde{V} (\mathbf{b}, \tau) \approx  \int d^2\mathbf{u} \, d\eta \widetilde{I} (\mathbf{u}, \eta) \theta_0^2  \,\widetilde{\gamma}(B_\textrm{chan} \eta) \int d\nu  \widetilde{A} \left[ \theta_0 \left( \mathbf{u} - \frac{\nu}{c} \mathbf{b} \right) \right] \phi \left( \frac{\nu - \nu_0}{B_\textrm{band}} \right) e^{i 2 \pi \nu (\eta - \tau)}.
\end{equation}
With our approximations, then, the effect of having frequency channels with non-zero width is to envelope our response in $\eta$: limited spectral resolution makes the array less sensitive to high $\eta$ modes (i.e., rapidly fluctuating spectral modes).

To further simplify our expression, it is useful to orient our $\mathbf{u} \equiv (u,v)$ axes so that the $u$ axis is in the same direction as the baseline vector $\mathbf{b}$.  If we further assume that the antenna's footprint on the $uv$ plane is separable, i.e.
\begin{equation}
\label{eq:SepPrimaryBeam}
\widetilde{A} \left[ \theta_0 \left( \mathbf{u} - \frac{\nu}{c} \mathbf{b} \right) \right] = \widetilde{A}_{b\parallel} \left[ \theta_0 \left( u - \frac{\nu}{c} b \right) \right] \widetilde{A}_{b\perp} (\theta_0 v),
\end{equation}
then our expression becomes
\begin{equation}
\label{eq:vIntegratedGenVtilde}
\widetilde{V} (\mathbf{b}, \tau) \approx  \int d^2\mathbf{u} \, d\eta \widetilde{I} (\mathbf{u}, \eta) \theta_0^2  \widetilde{A}_{b\perp} (\theta_0 v) \,\widetilde{\gamma}(B_\textrm{chan} \eta) \int d\nu \widetilde{A}_{b\parallel} \left[ \theta_0 \left( u - \frac{\nu}{c} b \right) \right] \phi \left( \frac{\nu - \nu_0}{B_\textrm{band}} \right) e^{i 2 \pi \nu (\eta - \tau)}.
\end{equation}
To proceed beyond this point requires specific forms for $\widetilde{A}_{b\parallel}$ and $\gamma$.  However, it is instructive to examine various limits.  For an instrument with short baselines and/or narrow fields of view satisfying $b\theta_0 \ll c/B_\textrm{band} $, $\widetilde{A}_{b\parallel}$ is a slowly varying function that can be factored out of the integral, yielding
\begin{equation}
\label{eq:shortBlVtilde}
\widetilde{V} (\mathbf{b}, \tau) \Bigg{|}_{b\theta_0 \ll c/B_\textrm{band}} \approx \int d^2\mathbf{u} \, d\eta \widetilde{I} (\mathbf{u}, \eta) \theta_0^2 B_\textrm{band} \widetilde{A} \left[ \theta_0 \left( \mathbf{u} - \frac{\nu_0}{c} \mathbf{b} \right) \right] \,\widetilde{\gamma}(B_\textrm{chan} \eta)  \widetilde{\phi} \left[ B_\textrm{band} (\eta - \tau) \right] e^{i 2\pi \nu_0 (\eta - \tau)},
\end{equation}
where we used  Eq. \eqref{eq:SepPrimaryBeam} to recombine $\widetilde{A}_{b\parallel}$ and $\widetilde{A}_{b\perp}$.  This gives the ``usual" description of an interferometric measurement: each baseline samples a portion of the sky in $uv\eta$ space, defined by the antenna's footprint on the $uv$ plane and the Fourier transform of the bandpass shape in $\eta$, enveloped by the Fourier transform of a frequency channel's profile.  On the other hand, for an instrument with long baselines and/or wide fields of view satisfying $b\theta_0 \gg c/B_\textrm{band} $, it is $\phi$ that is broad compared to $\widetilde{A}_{b\parallel}$, and  Eq. \eqref{eq:vIntegratedGenVtilde} is well-approximated by\footnote{As one factors $\phi$ out of the integral, the function can no longer enforce the limits on the $\nu$-integral that reflect the finite bandwidth of the instrument.  However, the $\widetilde{A}$ function that remains inside the integral is by construction narrow.  It is therefore perfectly acceptable to keep $-\infty$ and $+\infty$ as the limits of integration.}
\begin{equation}
\label{eq:longBlVtilde}
\widetilde{V} (\mathbf{b}, \tau) \Bigg{|}_{b\theta_0 \gg c/B_\textrm{band}} \approx \int d^2\mathbf{u} \, d\eta \widetilde{I} (\mathbf{u}, \eta) \frac{c \theta_0}{b} \widetilde{A}_{b\perp} (\theta_0 v) \,\widetilde{\gamma}(B_\textrm{chan} \eta)  \phi \left( \frac{uc/b - \nu_0}{B_\textrm{band}} \right) A_{b\parallel} \left[ \frac{c}{b \theta_0} ( \eta - \tau ) \right] e^{i 2\pi (\eta - \tau) \frac{uc}{b} }.
\end{equation}
\end{widetext}
We see that in this limit, the bandpass shape $\phi$ and the primary beam $A_{b\parallel}$ swap roles: the bandpass acts as the convolution kernel on the $uv$ plane, while the primary beam acts as the convolution kernel in the $\eta$-direction.  When $b\theta \sim c / B_\textrm{band}$, the full expression given by  Eq. \eqref{eq:vIntegratedGenVtilde} interpolates between the two extremes given by  Eqs. \eqref{eq:shortBlVtilde} and \eqref{eq:longBlVtilde}; in general, the bandpass shape and the primary beam both play a part in determining the $uv$ plane and $\eta$-direction convolution kernels.  This point was emphasized in Ref. \cite{Parsons2012b}.

Returning to the long baseline limit of  Eq. \eqref{eq:longBlVtilde}, we can see our first glimpse of the foreground wedge.  Suppose one were dealing with flat-spectrum foregrounds, where $\widetilde{I}(u,\eta) = \widetilde{I}_0 (u) \delta(\eta)$, so that there is no signal beyond $\eta=0$.  With such a sky, the measurement becomes proportional to $A_{b\parallel} ( c \tau / b \theta_0 ) $ and $| \widetilde{V} (\mathbf{b}, \tau) |^2 \propto A_{b\parallel}^2 ( c \tau / b \theta_0 )$.  Suppose we now make use of the delay approximation \cite{Parsons2012b}, where the quantity $| \widetilde{V} (\mathbf{b}, \tau) |^2$ can be treated as an estimate of the power spectrum $P(u,\eta)$ at $u \approx \nu_0 b/c$ and $\eta \approx \tau$.  The result is
\begin{equation}
\label{eq:firstGlimpseWedge}
P(u,\eta) \propto A_{b\parallel}^2 ( \nu_0 \eta / u \theta_0 ).
\end{equation}
Thus, even flat spectrum sources (which would naively only have power at $\eta=0$) gives a non-zero measured power spectrum at higher $\eta$.  If the primary beam is zero beyond some argument value (which is always true in some sense, since $A$ is identically zero below the horizon), then the power extends only to a finite region on the $u\eta$-plane.  For example, if we define $\theta_0$ to be the angle at which $A$ drops to zero (we are free to do so, since $\theta_0$ is simply a characteristic scale that we have so far kept general as simply ``some characteristic scale"),  Eq. \eqref{eq:firstGlimpseWedge} predicts that the line
\begin{equation}
\eta = \frac{\theta_0}{\nu_0} u
\end{equation}
should be a sharp boundary between zero and non-zero power.  Switching from angular/spectral Fourier coordinates $u$ and $\eta$ to comoving spatial wavenumbers $k_\perp$ and $k_\parallel$ using the relations in Appendix \ref{appendix:Fourier} gives
\begin{equation}
\label{eq:CosmologicalCoordsWedge}
k_\parallel = \frac{H_0 D_c E(z) \theta_0}{c (1+z)} k_\perp,
\end{equation}
where $E(z) \equiv \sqrt{\Omega_m (1+z)^3 + \Omega_\Lambda}$, $D_c$ is co-moving line-of-sight distance, $H_0$ is the Hubble parameter, $\Omega_m$ is the normalized matter density, and $\Omega_\Lambda$ is the normalized dark energy density.  This is precisely the ``usual" formula given for the edge of the foreground wedge (e.g., in Refs. \cite{Morales2012,Trott2012,Parsons2012b,Dillon2014}).  As a function of $\eta$ (or $k_\parallel$), the wedge has a profile given by the square of the primary beam profile, scaled by $u$ in such a way that power is found at higher $\eta$ for higher $u$.

In presenting a preview of the wedge, we have made a number of assumptions that will be relaxed in the following sections.  For example, we will no longer approximate $\tau$ modes as $\eta$ modes.  Nor will we assume that each baseline cleanly samples just a single value of $u$.  In fact, our use of the delay approximation here was somewhat inappropriate---while one may always take a delay \emph{transform}, we have seen (e.g., from Figure \ref{fig:ChromaticBaselines}) that the delay \emph{approximation} (i.e., assuming $\tau\sim \eta$) ought to work well only if baselines are short.  By using Eq. \eqref{eq:longBlVtilde} instead of Eq. \eqref{eq:shortBlVtilde}, however, we are expressly working in the long-baseline limit.  It is thus important to emphasize that the preview shown here is included only to build intuition, and a much more rigorous treatment will be presented in later sections.  There, we will also generalize to a sky with an arbitrary power spectrum.  While there will be minor alterations to details of the foreground wedge, the basic picture will remain intact.

In the strictest sense, the derivation that we have just presented is nothing new.  We have simply re-derived a number of results (e.g. the existence of the wedge) that are already known in the literature.  However, our re-derivations have been part of an analytic formalism, confirming a number of numerical results while bringing together many known (but previously separate) features in a unified framework.  In the next section, we will extend this framework to include a fully covariant description of the wedge, including correlated errors in such a way that extends the quadratic power spectrum estimation techniques of Ref. \cite{Liu2011} to include wedge physics.  Setting up such a framework allows one to systematically examine the statistical properties of various power spectrum estimators in light of the wedge, a study that we perform in Paper II.

\section{Quadratic estimator formalism}
\label{sec:QuadEst}
Until now, our focus has been on the measurement of a visibility, which is \emph{linear} in temperature.  The power spectrum, however, is a quantity that depends quadratically on temperature.  In this section we very briefly review the mathematical machinery that makes possible the fully covariant description of the EoR window we will present in Sections \ref{sec:basicEstAnalytic} and \ref{sec:basicEstNumerical}.  The basis of our discussion will be the quadratic estimator formalism, which has a long history in the CMB and galaxy survey literature (e.g., Refs. \cite{Tegmark1997b,Bond1998,Tegmark1998}), and was explicitly adapted for $21\,\textrm{cm}$ cosmology in Refs. \cite{Liu2011,Dillon2013}.

The central quantity that encodes our instrument (and therefore the statistical properties of our estimators) is the data covariance matrix $\mathbf{C}$.  To form the covariance matrix, imagine that our input data (organized by baseline and delay mode) is serialized into a data vector $\mathbf{x}$, i.e.,
\begin{equation}
\label{eq:xvectdef}
\mathbf{x} = \left( \begin{array}{c}
\widetilde{V} (\mathbf{b}_1, \tau_1) \\
\widetilde{V} (\mathbf{b}_1, \tau_2) \\
\vdots \\
\widetilde{V} (\mathbf{b}_2, \tau_1) \\
\widetilde{V} (\mathbf{b}_2, \tau_2) \\
\vdots
\end{array}
\right).
\end{equation}
The covariance matrix is then given by $\mathbf{C} \equiv \langle \mathbf{x} \mathbf{x}^\dagger \rangle$.  Importantly, we will keep track of all off-diagonal elements in the matrix, so that all correlations between different baselines and different delay (or spectral) modes are taken into account.  Knowledge of these correlations will allow us to formulate both a covariant description of the foreground wedge and the tools to fight its contaminating influence.

Although we omitted noise contributions in the previous section for notational cleanliness, they of course contribute to the variance captured by $\mathbf{C}$.  Assuming that instrumental noise is uncorrelated with sky signals, the noise appears as an additive term to the sky covariance, so that we can write $\mathbf{C}$ as
\begin{equation}
\label{eq:C=N+S}
\mathbf{C} \equiv \mathbf{N} + \mathbf{S},
\end{equation}
where $\mathbf{S}$ is the sky signal portion of the covariance.  Inserting  Eq. \eqref{eq:genVtilde} into our general definition of the data covariance, we obtain
\begin{equation}
\label{eq:genCovar}
\mathbf{S}_{ij} =\int d^2 \mathbf{u} d\eta \, P (\mathbf{u}, \eta) h(\mathbf{u} , \eta; \mathbf{b}_i, \tau_i) h^*(\mathbf{u} , \eta; \mathbf{b}_j, \tau_j)
\end{equation}
where we have defined
\begin{eqnarray}
\label{eq:hDef}
h(\mathbf{u}, \eta; &&\mathbf{b}, \tau)  \equiv \theta_0^2 \widetilde{\gamma} ( B_\textrm{chan} \eta ) \times \nonumber \\
&& \int d\nu  \widetilde{A} \left[ \theta_0 \left( \mathbf{u} - \frac{\nu}{c} \mathbf{b} \right) \right] \phi \left( \frac{\nu - \nu_0}{B_\textrm{band}} \right) e^{i 2 \pi \nu (\eta - \tau)}, \qquad
\end{eqnarray}
and have made use of the definition of the power spectrum $P (\mathbf{u}, \eta)$:
\begin{equation}
\langle \widetilde{I} (\mathbf{u}, \eta) \widetilde{I}^* (\mathbf{u}^\prime, \eta^\prime) \rangle \equiv P(\mathbf{u}, \eta) \delta(\mathbf{u} - \mathbf{u}^\prime) \delta(\eta - \eta^\prime).
\end{equation}
As we have written it, our power spectrum is expressed in terms of $\mathbf{u}$ and $\eta$, instead of the more common combination of $\mathbf{k}_\perp$ and $k_\parallel$.  Doing so minimizes the number of cosmological quantities in our expressions, since $\mathbf{u}$ and $\eta$ are the more ``natural" quantities from the perspective of the instrument.  This represents no loss of generality, and indeed, in Section \ref{sec:basicEstNumerical} we will express our results in terms of $\mathbf{k}_\perp$ and $k_\parallel$.

To form a practical estimator for the power spectrum, it is necessary to discretize.  Assuming that the power spectrum possesses cylindrical symmetry\footnote{The true power spectrum of the cosmological signal is of course spherically symmetric (thanks to statistical isotropy), and thus can be further binned.  However, systematics such as foregrounds tend to be cylindrically symmetric, since the instrument probes the line-of-sight direction differently than it does the angular directions \cite{Morales2004}.  Another effect to consider (though it is beyond the scope of this paper) is that of redshift-space distortions, which will break spherical symmetry.  The cylindrical power spectrum is therefore a useful diagnostic quantity to compute prior to the formation of a spherical power spectrum.  Note that in general, we need not make any assumptions about symmetry in our estimation formalism.  If desired, our formalism can be used to estimate $P(\mathbf{u}, \eta)$ without even the cylindrical binning step.  The presence of symmetry, while useful for increasing signal-to-noise, is much less important than other assumptions such as the smoothness of foregrounds.  (Although see Section \ref{sec:Unsmooth} for a discussion of strategies for dealing with unsmooth foregrounds).} (so that the power depends only on the magnitudes of $\mathbf{u}$ and $\eta$), one can imagine binning the $uv\eta$ space into a series of annuli, each specified by a radius $|u| =\sqrt{u^2 + v^2}$ and ``vertical distance" $|\eta|$ away from the $\eta=0$ plane.  Each annulus can then be represented as a small cell on a $|u|\eta$ plane (which we will henceforth call the $u\eta$ plane to conform to convention).\footnote{We note an unfortunate bit of notation: $u$ is used both to denote the first coordinate on the $uv$ plane as well as the magnitude of the $\mathbf{u}$ vector.  Unfortunately, both usages are standard.}  As long as these cells are made sufficiently small, the power spectrum can be approximated by a constant \emph{bandpower} $p_\alpha$ in each cell, with $\alpha$ indexing a serialized list of locations on the $u\eta$ plane.  With this approximation, our covariance can be written compactly as
\begin{equation}
\label{eq:CovarDecomp}
\mathbf{C} = \mathbf{N} + \sum_\alpha p_\alpha \mathbf{C}_{,\alpha}
\end{equation}
where $\mathbf{C}_{,\alpha} \equiv \partial \mathbf{C} / \partial p_\alpha$ is the response of the covariance to $\alpha$th bandpower, and is given by
\begin{equation}
\label{eq:CcommaAlpha}
(\mathbf{C}_{,\alpha})_{ij} =  \int_{V_\alpha} d^2 \mathbf{u} d\eta \,  h(\mathbf{u} , \eta; \mathbf{b}_i, \tau_i) h^*(\mathbf{u} , \eta; \mathbf{b}_j, \tau_j),
\end{equation}
with $V_\alpha$ denoting the annular volume that is binned into the $\alpha$th $u\eta$ cell.  If desired, the sky contribution of the covariance can be further divided into separate contributions from foregrounds and the cosmological signal:
\begin{equation}
\label{eq:genCovarDecomp}
\mathbf{C} = \mathbf{N} + \mathbf{C}_\textrm{fg} +  \sum_\alpha p_\alpha^\textrm{sig} \mathbf{C}_{,\alpha},
\end{equation}
where $\mathbf{C}_\textrm{fg}$ is the foreground covariance, and $p_\alpha^\textrm{sig}$ is the $\alpha$th bandpower of the cosmological signal only.   Eq. \eqref{eq:genCovarDecomp} is more general than  Eq. \eqref{eq:CovarDecomp}, because the latter implicitly assumes that $\mathbf{C}_\textrm{fg}$ is given by $\sum_\alpha p_\alpha^\textrm{fg} \mathbf{C}_{,\alpha}$, where $p_\alpha^\textrm{fg}$ is a set of foreground bandpowers.  This assumption holds true only when the foregrounds are describable as a power spectrum.

In the quadratic estimator formalism, the bandpowers are extracted by forming weighted pairwise combinations of the data vector $\mathbf{x}$.  In particular, one can form a quadratic estimator $\widehat{p}_\alpha$ of the true bandpower $p_\alpha$ by computing\footnote{Throughout this paper, we use hats to denote estimators of quantities.}
\begin{equation}
\label{eq:GenQuadEst}
\widehat{p}_\alpha = \mathbf{x}^\dagger \mathbf{E}^\alpha \mathbf{x},
\end{equation}
where $\mathbf{E}^\alpha$ is an estimator matrix of weights to be used for weighting pairwise products of the data when estimating the $\alpha$th bandpower.  To see how this works, consider (as a toy example) a noiseless cosmological survey 
with uncorrelated real-space measurements in a three-dimensional volume.  Further suppose that one's goal is to measure the unbinned, three-dimensional power spectrum $P(\mathbf{k})$, i.e., $\widehat{p}_\alpha = \widehat{P}(\mathbf{k}_\alpha)$.  If $\mathbf{x}$ is expressed in a real-space basis (so that it is simply a serialized list of real-space voxel intensities), a sensible choice for the estimator matrix would be $\mathbf{E}^\alpha_{ij} \propto e^{-i \mathbf{k}_\alpha \cdot (\mathbf{r}_i - \mathbf{r}_j)}$, where $\mathbf{r}_i$ and $\mathbf{r}_j$ are the position vectors of the $i$th and $j$th voxels, respectively.  We therefore see that in this example, the role of $\mathbf{E}^\alpha$ is to take a Fourier transform of the data.  If one desires estimates of a \emph{binned} power spectrum (for example, one where statistical isotropy allows the binning of power over shells of constant $|\mathbf{k}|$), the relevant $\mathbf{E}^\alpha$ for different $\mathbf{k}_\alpha$ are simply averaged together in each bin.

Now suppose instead that our data are expressed in a Fourier basis, so that each element of the data vector $\mathbf{x}$ represents the Fourier amplitude at some location in $\mathbf{k}$ space.  The estimator matrix is then even simpler, and in fact becomes diagonal, with $\mathbf{E}^\alpha_{ij} = \delta_{\alpha i} \delta_{ij}$.  In our current application, the data are organized by baseline $\mathbf{b}$ and delay $\tau$.  As discussed above, $\mathbf{b}$ and $\tau$ closely approximate the Fourier wavenumbers $\mathbf{u}$ and $\eta$ in some regimes, but the correspondence is not perfect.  For our application we would therefore expect $\mathbf{E}^\alpha$ to be diagonal-dominant, but not be perfectly diagonal.  (For an explicit form, see Sections \ref{sec:basicEstAnalytic} and \ref{sec:basicEstNumerical}).

In general, correlated errors and other instrumental effects (such as the ones that we seek to model in this paper) make the estimator matrices more complicated than they were in our pedagogical examples.  They will typically involve the $\mathbf{C}_{,\alpha}$ matrices, since those provide the link between the ``input" space that the data covariance inhabits, and the ``output" space of bandpowers.  The detailed form of the family of $\mathbf{E}_\alpha$ matrices is a choice made by the data analyst, and different choices yield estimators with different statistical properties.  One such property is the error covariance $\boldsymbol \Sigma_{\alpha \beta} \equiv \langle \widehat{p}_\alpha \widehat{p}_\beta \rangle - \langle \widehat{p}_\alpha \rangle \langle \widehat{p}_\beta \rangle $ of our estimated bandpowers, which is given by\footnote{While we write all vector and matrix quantities in boldface, it is important to note that there are two different vector spaces at work here.  The error covariance $\boldsymbol \Sigma$ and the window function matrix $\mathbf{W}$ defined later inhabit the ``output" vector space indexed by locations on the $u\eta$ plane, unlike matrices such as $\mathbf{N}$ and $\mathbf{C}$, which inhabit the ``input" vector space indexed by baseline and delay.  Hybrid quantities include $\mathbf{C}_{,\alpha}$ and $\mathbf{E}^\alpha$, which can either be thought of as a family of matrices in the input vector space, or as rank-3 tensors in a combined space.}
\begin{equation}
\label{eq:matrixErrorCovar}
\boldsymbol \Sigma_{\alpha \beta} = 2 \textrm{tr} \left[ \mathbf{C} \mathbf{E}^\alpha \mathbf{C} \mathbf{E}^\beta \right],
\end{equation}
a result that can be derived by direct substitution of  Eq. \eqref{eq:GenQuadEst} into the definition of the error covariance.  The error bars on our bandpower estimates are given by $\Delta p_\alpha \equiv (\boldsymbol \Sigma_{\alpha \alpha})^{\frac{1}{2}}$, but it is important to note that the error covariance contains much more information than just the error bars: off-diagonal elements of the covariance encode correlations between different $u\eta$ cells of the cylindrical power spectra.  With a fully covariant formulation of the errors, it is possible to over-resolve in $u\eta$ space, evading the commonly-made assumption that $u\eta$ cells are independent (i.e., have a diagonal covariance matrix) so long as they are more than $1/B_{band}$ apart in the $\eta$ direction and separated by more than the width of $\widetilde{A}$ in the $u$ direction.  While this assumption is prevalent in the $21\,\textrm{cm}$ cosmology literature for reasons of computational simplicity, we will see that it is one that should be avoided.  More precisely, we will see in Section \ref{sec:basicEstNumerical} that the errors on the $u\eta$ plane are not independent at high $k_\perp$ (including the foreground wedge region), necessitating a full accounting of the entire error covariance matrix, and not just its diagonal elements.

In addition to the error covariance, the quadratic estimator formalism also allows the computation of biases and window functions.  Taking the expectation value of  Eq. \eqref{eq:GenQuadEst}, recalling that $\mathbf{C} \equiv \langle \mathbf{x} \mathbf{x}^\dagger \rangle$, and inserting  Eq. \eqref{eq:genCovarDecomp} gives
\begin{eqnarray}
\widehat{p}_\alpha &=& \sum_\beta \textrm{tr} [\mathbf{E}^\alpha \mathbf{C}_{,\beta} ] \,p_\beta^\textrm{sig} + \textrm{tr}  [\mathbf{E}^\alpha (\mathbf{N}+ \mathbf{C}_\textrm{fg})] \nonumber \\
&\equiv& \sum_\beta \mathbf{W}_{\alpha \beta} \,p_\beta^\textrm{sig} + b_\alpha, \label{eq:EstWindAndBias}
\end{eqnarray}
where we have defined the window function matrix
\begin{equation}
\label{eq:WindGeneral}
\mathbf{W}_{\alpha \beta} \equiv \textrm{tr} [\mathbf{E}^\alpha \mathbf{C}_{,\beta} ] 
\end{equation}
and the contamination bias
\begin{equation}
\label{eq:bias}
b_\alpha = \textrm{tr}  [\mathbf{E}^\alpha (\mathbf{N}+ \mathbf{C}_\textrm{fg})] .
\end{equation}
From  Eq. \eqref{eq:EstWindAndBias}, we see that each row of the window function matrix gives a \emph{window function} that specifies the linear combination of the true bandpowers that each estimate of a bandpower represents.  Typically, the $\mathbf{E}^\alpha$ matrix is normalized such that each row of $\mathbf{W}$ sums to unity, allowing the linear combinations to be interpreted as weighted averages.\footnote{In the signal processing literature, the tapering function $\phi$ that we introduced in the previous section is often called a ``window function".  This is a conceptually-separate use of the term, and we avoid the signal processing nomenclature in order to be consistent with the cosmology literature.  For us, a window function will always refer to the linear combination of the true power spectrum that forms a particular bandpower estimate, and never the tapering function.}

The contamination bias represents an additive bias to the estimated power spectrum that arises due to residual noise and foregrounds in the data.  In practice, cross-correlation techniques---such as forming cross-power spectra between odd and even time samples of data, as was done in Ref. \cite{Dillon2014}, or between different subsets of redundant baselines in an array, as was done in Ref. \cite{Parsons2013}---allow the noise bias to be eliminated without any explicit bias subtraction.  The bias that one needs contend with is therefore solely comprised of the foreground bias:
\begin{equation}
\label{eq:fgBias}
b_\alpha = b^\textrm{fg}_\alpha = \textrm{tr}  [\mathbf{E}^\alpha  \mathbf{C}_\textrm{fg}] .
\end{equation}
If a perfect foreground model is available, this expected level of this bias can be computed and subtracted from the power spectrum estimate.  However, because a detailed knowledge of the low-frequency sky is as-yet unavailable, this subtraction step is often omitted to avoid over-subtractions that destroy cosmological information.  Instead, one simply hopes that the bias is small in regions of the $u\eta$ plane where one wishes to make a power spectrum measurement.

In the following sections, the error covariance $\boldsymbol \Sigma$, the window function matrix $\mathbf{W}$, and the bias $b_\alpha$ are the quantities that will provide us with a detailed, covariant picture of the EoR window and the foreground wedge.  The bias essentially captures the power spectrum contribution from noise and foreground contaminants, and corresponds to the foreground wedge signatures seen in various simulations in the literature.  The window functions provide an alternate view of the wedge: the wedge can be thought of as a leakage of power from low to high $\eta$ (or equivalently, $k_\parallel$) modes that becomes increasingly pronounced as $u$ (or $k_\perp$) increases.  For a wedge to exist, window functions for bandpowers centered at high $\eta$ and high $u$ must have tails that extend to low $\eta$, where foregrounds live.  Finally, the error covariance provides an estimate of the error bars throughout the Fourier plane (including the wedge region), as well as a quantification of how the chromatic response of an interferometer can cause error correlations between otherwise uncorrelated $u\eta$ cells.

\section{A covariant description of the foreground wedge for a basic estimator}
\label{sec:basicEstAnalytic}

In this section, we use the quadratic estimator formalism to derive the foreground wedge and the EoR window for a ``basic" estimator of the power spectrum.  We will make a number of approximations for the sake of analytical tractability, leaving an exact numerical treatment to Section \ref{sec:basicEstNumerical}.  The goal here is to formalize the discussion from Section \ref{sec:InstrumentResponse} to obtain a fully covariant, analytic description of power spectrum properties at high $u$ and low $\eta$, where the foreground wedge resides.  This will provide a basic picture of the foreground challenges that we face, setting the stage for Paper II, where we look at how these challenges can be mitigated with better estimators.

The relatively simple estimator that we will examine in this paper is specified by the relation
\begin{equation}
\label{eq:basicEstEalpha}
\mathbf{E}^\alpha = M_\alpha\, \mathbf{N}^{-1} \mathbf{C}_{,\alpha} \mathbf{N}^{-1},
\end{equation}
where $\mathbf{N}$ is the instrumental noise covariance and $M_\alpha$ is a normalizing scalar\footnote{In this paper, we do not consider the more general possibility of a matrix-based normalization, where instead of a simple multiplicative normalization, one multiplies the unnormalized bandpower estimates $\hat{p}_\beta^\prime$ by a matrix $\mathbf{M}$ to form the normalized bandpowers $\hat{p}_\alpha$, i.e., $\hat{p}_\alpha = \sum_\beta \mathbf{M}_{\alpha \beta} \hat{p}_\beta^\prime$.  For details, please see Ref. \cite{Dillon2014} or Paper II.} for each bandpower $\alpha$.  This choice gives an estimator that is quite similar to the crude $| \widetilde{V} (\mathbf{b}, \tau) |^2$ estimator discussed informally in Section \ref{sec:InstrumentResponse}.  The principal difference between what we will consider here and our previous estimator is the presence of $\mathbf{C}_{,\alpha}$.  The role of $\mathbf{C}_{,\alpha}$ is two-fold.  Its first purpose is to complete a signal-to-noise weighting of our data: the copies of $\mathbf{N}^{-1}$ downweight noisy data, while $\mathbf{C}_{,\alpha}$ (by virtue of its being the derivative of $\mathbf{C}$) upweights the high signal portions.  The second purpose is to map the data from the input baseline $\mathbf{b}$ and delay $\tau$ space to the output $u\eta$ space.  Recall from Section \ref{sec:InstrumentResponse} that while $\mathbf{b}$ and $\tau$ approximate $\mathbf{u}$ and $\eta$, respectively, the correspondence is not perfect.  Applying $\mathbf{C}_{,\alpha}$ completes the transition to Fourier space, a fact that will become more apparent when we write down an explicit form for the matrix.

Studying the basic estimator given by Eq. \eqref{eq:basicEstEalpha} is worthwhile because it is approximately equivalent to the methods used in a number of state-of-the-art $21\,\textrm{cm}$ power spectrum pipelines for  analyzing observations and simulations \cite{Bernardi2013,Thyagarajan2013,Hazelton2013}.  These pipelines typically use an optimal mapmaking approach \cite{Tegmark1997a,Morales2008} to first go from visibilities to a gridded $uv\eta$ data cube of Fourier amplitudes.  The complex magnitudes of these amplitudes are then squared and binned to estimate power spectrum bandpowers.  (Note that while the mapmaking may be optimal in this case, the subsequent power spectrum estimation is not).  In Appendix \ref{appendix:GridVsVisEquiv}, we will prove that in the limit of infinitely fine bins in Fourier space, such pipelines are equivalent to estimating power spectra directly from the visibilities using Eq. \eqref{eq:basicEstEalpha} and the quadratic estimator formalism.  Our numerical results will therefore be roughly representative of those seen in the aforementioned pipelines, but with fewer gridding artifacts because we go straight from visibilities to power spectra.

For analytical tractability here and numerical tractability in later sections, we will use an approximate form for the covariance matrix:
\begin{equation}
\mathbf{C}_{ij} = \int du d\eta P(u,\eta) g(u, \eta; b_i, \tau_i) g^* (u, \eta; b_j, \tau_j)
\end{equation}
where
\begin{eqnarray}
\label{eq:GeeDef}
g &&(u, \eta; b_i,  \tau_i) \equiv   k \, \widetilde{\gamma}(B_\textrm{chan} \eta) \times \nonumber \\ 
&&  \int d\nu  \widetilde{A}_{b\parallel} \left[ \theta_0 \left( u - \frac{\nu}{c} b_i \right) \right] 
\phi \left( \frac{\nu - \nu_0}{B_\textrm{band}} \right) e^{i 2 \pi \nu (\eta - \tau_i)}
\end{eqnarray}
with
\begin{equation}
k \equiv \theta_0 \left( \int dq \widetilde{A}_{b\perp}^2 (\theta_0 q) \right)^\frac{1}{2},
\end{equation}
and we have once again omitted the instrumental noise contribution to the covariance for simplicity.  Superficially, this looks quite similar to  Eqs. \eqref{eq:genCovar} and \eqref{eq:hDef}.  However, in this case we have gone beyond simply forming $\mathbf{C} \equiv \langle \mathbf{x} \mathbf{x}^\dagger \rangle$ from  Eq. \eqref{eq:vIntegratedGenVtilde}, in that we have performed the integral over $\widetilde{A}_{b\perp}$.  This is not always permissible, and represents a subtle additional approximation: we have assumed that baselines that are similar in length but very different in orientation have a negligible correlation with each other, and that those with similar orientations are correlated as though they were identical in orientation.  In other words, we assume that although two baselines can be completely uncorrelated, partially redundant, or perfectly redundant in the direction of the baseline vector,\footnote{Note that our ability to accommodate the full continuum of redundancy from zero to perfect redundancy along the direction of the baseline vector allows us to capture possible subtleties related to partial redundancy effects like those highlighted in Ref. \cite{Hazelton2013}.} overlaps between baselines in the transverse direction are treated in a binary fashion, so that the overlap is either zero or perfect.  This assumption was inherited from the derivation of  Eq. \eqref{eq:vIntegratedGenVtilde}, which required a re-orientation of the axes of the $uv$ plane so that the $u$ axis would lie along the direction of the baseline.  While this can always be done for a single baseline, the covariance matrix $\mathbf{C}$ encodes correlations between different baselines, which may be oriented differently.  It is thus strictly speaking incorrect to form a covariance matrix from  Eq. \eqref{eq:vIntegratedGenVtilde}, and in principle one should use  Eq. \eqref{eq:genVtilde} instead.  For the purposes of intuition, however, we may continue with our approximate expression as long as we remember that distant baselines have negligible correlation.

As we previously mentioned, $\mathbf{C}_{,\alpha}$ provides the crucial link between the input data and the output Fourier space.  It therefore forms a crucial component of any error statistic.  Working in the limit of a continuous (rather than discrete) set of bandpowers, we may differentiate $\mathbf{C}$ with respect to $P(u_\alpha, \eta_\alpha)$ to obtain
\begin{equation}
\label{eq:CcommaAlphaggstar}
(\mathbf{C}_{,\alpha})_{ij} \equiv \frac{\partial \mathbf{C}_{ij}}{\partial P(u_\alpha, \eta_\alpha)} = g(u_\alpha, \eta_\alpha; b_i, \tau_i) g^* (u_\alpha, \eta_\alpha; b_j, \tau_j),
\end{equation}
where we used the fact that $P(u, \eta)$ can be written as $\int du_\alpha d\eta_\alpha P(u_\alpha, \eta_\alpha) \delta(u - u_\alpha) \delta (\eta - \eta_\alpha) $.  Inserting this into Eq. \eqref{eq:basicEstEalpha} provides a concrete example of the general proof of equivalence given in Appendix \ref{appendix:GridVsVisEquiv}.  One sees that each copy of $g$ acts on a noise-weighted copy of the data vector $\mathbf{x}$.  Examining Eq. \eqref{eq:GeeDef} reveals that the action of $g$ is to Fourier transform the delay spectrum back into the frequency domain, apply another copy of the tapering functions, grid the result at the appropriate location on the $uv$ plane, and then Fourier transform in frequency again, before adjusting for frequency channel discretization by an additional weighting in $\eta$.  This is precisely the procedure that one would follow with a mapmaking algorithm in $uv\eta$ space \cite{Morales2008}.  The result is then squared to form a power spectrum.  While this particular example may at first sight seem to render the delay basis obsolete (since the first action of $g$ is to transform back to a frequency spectrum), it is important to remember that in a more realistic case, one may be unable approximate the bandpowers as being continuous.  For example, at small $u$ and $\eta$, bin sizes may be comparable to the values of $u$ and $\eta$ themselves.  Many of the algebraic simplifications used in this section then become inapplicable, necessitating full numerical manipulations of the relevant matrices, which are typically more computationally efficient in the delay basis (as we discuss in Appendix \ref{appendix:Computational}).

Continuing with our approximation scheme for this section, however, Eq. \eqref{eq:CcommaAlphaggstar} is particularly convenient for computing our suite of error statistics because it is separable.  Taking advantage of this, the window functions for our basic estimator reduce to
\begin{eqnarray}
\mathbf{W}_{\alpha \beta} &=& \textrm{tr} [ \mathbf{E}^\alpha \mathbf{C}_{,\beta} ] \propto  \textrm{tr} [\mathbf{C}_{,\alpha}  \mathbf{C}_{,\beta}] = | \mathbf{g}^{\alpha \dagger} \cdot \mathbf{g}^{\beta}|^2 \nonumber \\ 
& \approx & \Bigg{|} \sum_{b} \int d\tau \, g^*(u_\alpha, \eta_\alpha; b, \tau) g (u_\beta, \eta_\beta; b, \tau) \Bigg{|}^2,\quad \label{eq:simplifiedWind}
\end{eqnarray}
where we defined the shorthand $\mathbf{g}^\alpha \equiv g(u_\alpha, \eta_\alpha; b, \tau)$, and in the last step assumed that our delay bins were fine enough to be approximated as being continuous.  This form for the window function matrix has a straightforward geometric interpretation: when estimating the $\alpha$th bandpower, one probes a mixture of the true bandpowers; the amount of the $\beta$th band that is included in the estimate of the $\alpha$th band is given by the overlap of our interferometer's response to the $\alpha$th and $\beta$th bands.  We stress, however, that this simple form does not hold when one considers more complicated estimators such as the ones that we will consider in Paper II.

Let us now compute some example window functions.  Just as we did in Section \ref{sec:InstrumentResponse}, we can gain some analytic intuition by working in the short and long baseline limits.  For short baselines satisfying $b \theta_0 \ll c / B_\textrm{band}$ (or equivalently, if the time-delay $\tau$ of a signal between the antennas of the baseline satisfies $\tau \ll 1/ \theta_0 B_\textrm{band}$), we may invoke the same approximations that led to Eq. \eqref{eq:shortBlVtilde}, and say that
\begin{eqnarray}
g^\textrm{short bl} (u, \eta; b, \tau) \propto \,\widetilde{\gamma}(B_\textrm{chan} \eta)  \widetilde{A}_{b\parallel} \left[ \theta_0 \left( u - \frac{\nu_0}{c} b \right) \right] \nonumber \\
\times \widetilde{\phi} \left[ B_\textrm{band} (\eta - \tau) \right] e^{i 2\pi \nu_0 (\eta - \tau)}. \quad
\label{eq:ShortBlGeeDef}
\end{eqnarray}
Inserting Eq. \eqref{eq:ShortBlGeeDef} into Eq. \eqref{eq:simplifiedWind} and evaluating the integral over $\tau$ yields
\begin{eqnarray}
\label{eq:ShortBlAnalyticalWindow}
&&W_{u_\alpha,\eta_\alpha}^\textrm{short bl} (u_\beta,\eta_\beta) \propto \widetilde{\gamma}^2(B_\textrm{chan} \eta_\alpha)\widetilde{\gamma}^2(B_\textrm{chan} \eta_\beta)  \nonumber \\
&& \times \left( \sum_b   \widetilde{A}_{b\parallel} \left[ \theta_0 \left( u_\alpha - \frac{\nu_0}{c} b \right) \right]  \widetilde{A}_{b\parallel} \left[ \theta_0 \left( u_\beta - \frac{\nu_0}{c} b \right) \right] \right)^2 \, \nonumber \\
&& \times \left(\widetilde{\phi^2} \left[B_\textrm{band} (\eta_\alpha - \eta_\beta) \right] \right)^2,
\end{eqnarray}
where $\widetilde{\phi^2}$ signifies the Fourier transform of $\phi^2$, not the square of $\widetilde{\phi}$.  This expression is in line with what one might intuitively expect from interferometry: the spatial $u$-dependence of the window function is controlled by the primary beam (or more precisely, its Fourier transform), while the spectral $\eta$-dependence is controlled by the bandpass.

On the other hand, with long baselines satisfying $b \theta_0 \gg c / B_\textrm{band}$ we can use the approximations that led to Eq. \eqref{eq:longBlVtilde}, obtaining
\begin{eqnarray}
g^\textrm{long bl}(u,\eta; b,\tau) \propto \widetilde{\gamma}(B_\textrm{chan} \eta) \phi \left( \frac{uc/b - \nu_0}{B_\textrm{band}} \right) \nonumber \\
\times  \int d\nu  \widetilde{A}_{b\parallel} \left[ \theta_0 \left( u - \frac{\nu}{c} b \right) \right]  e^{i 2\pi \nu (\eta - \tau)}.
\end{eqnarray}
Once again, we may insert this into Eq. \eqref{eq:simplifiedWind} to get
\begin{widetext}
\begin{eqnarray}
\label{eq:LongBlSomeLimit}
W_{u_\alpha,\eta_\alpha}^\textrm{long bl} (u_\beta,\eta_\beta) \propto \widetilde{\gamma}^2(B_\textrm{chan} \eta_\alpha)\widetilde{\gamma}^2(B_\textrm{chan} \eta_\beta) \Bigg{|} \sum_b \phi \left( \frac{u_\alpha c / b-\nu_0}{B_\textrm{band}}\right) \phi \left( \frac{u_\beta c / b-\nu_0}{B_\textrm{band}}\right) \nonumber \\
 \int d\nu \widetilde{A}^*_{b\parallel} \left[ \theta_0 \left( u_\alpha - \frac{\nu}{c} b \right) \right]  \widetilde{A}_{b\parallel} \left[ \theta_0 \left( u_\beta - \frac{\nu}{c} b \right) \right] e^{-i 2 \pi \nu (\eta_\alpha - \eta_\beta)} \Bigg{|}^2,
\end{eqnarray}
Now, consider first the $\eta$-dependence (rather than the $u$-dependence) of the window functions, since our principal worry is that smooth, low-$\eta$ foregrounds might scatter to higher $\eta$ because of the instrument's chromaticity.  If we suppress the $u$ dependence by setting $u = u_\alpha = u_\beta$, we obtain
\begin{equation}
\label{eq:WedgeWindow}
W_{\eta_\alpha}^\textrm{long bl} (\eta_\beta) \propto \widetilde{\gamma}^2(B_\textrm{chan} \eta_\alpha)\widetilde{\gamma}^2(B_\textrm{chan} \eta_\beta) \Bigg{|} \sum_b \frac{c}{b} e^{i 2\pi \frac{uc}{b} (\eta_\beta-\eta_\alpha)} \phi^2 \left( \frac{\frac{u c}{b}-\nu_0}{B_\textrm{band}}\right) (A_{b\parallel} \ast A_{b\parallel} ) \left[ \frac{c}{b\theta_0} (\eta_\beta - \eta_\alpha) \right] \Bigg{|}^2.
\end{equation}
\end{widetext}

 Eq. \eqref{eq:WedgeWindow} predicts that with our basic estimator, foregrounds should appear in the now-familiar wedge in $u\eta$-space.  To see this, consider the additive foreground bias for a hypothetical single-baseline interferometer.  Since we are concerned with foregrounds, the most relevant regions of the $u\eta$ plane will be the low $\eta$ regions.  We may therefore safely ignore the $\widetilde{\gamma}^2$ terms, since at low $\eta$ they will be approximately unity anyway.  If we imagine that such foregrounds are described by a power spectrum $p_\alpha^\textrm{fg}$, the foreground covariance is given by $\sum_\alpha p_\alpha^\textrm{fg} \mathbf{C}_{,\alpha}$, and the foreground contribution $b_\alpha^\textrm{fg}$ to the bias in  Eq. \eqref{eq:bias} is given by
\begin{subequations}
\begin{eqnarray}
\label{eq:WindToBias}
b_\alpha^\textrm{fg} &=& \textrm{tr}  [\mathbf{E}^\alpha\mathbf{C}_\textrm{fg}] = \sum_\beta  \textrm{tr} [ \mathbf{E}^\alpha \mathbf{C}_{,\beta} ]  p_\beta^\textrm{fg} = \sum_\beta \mathbf{W}_{\alpha \beta} p_\beta^\textrm{fg} \\
& \propto &\!\! \int \! \! \left( ( A_{b\parallel} \ast A_{b\parallel}) \left[ \frac{c}{b \theta_0} (\eta_\alpha - \eta_\beta) \right]\right)^2 \!\!\! P^{\textrm{fg}} (\eta_\beta)  d\eta_\beta, \qquad
\end{eqnarray}
\end{subequations}
where we have suppressed the $u$ dependence of the foreground power spectrum for notational cleanliness.  Now, if we approximate the foregrounds as being completely comprised of flat spectrum sources, then $P^\textrm{fg} (\eta) \propto \delta(\eta)$, and our foreground bias becomes
\begin{equation}
\label{eq:anotherWedge}
b^\textrm{fg} (\eta) \propto \left( ( A_{b\parallel} \ast A_{b\parallel}) \left[ \frac{c}{b \theta_0} \eta  \right]\right)^2.
\end{equation}
This equation provides a general mathematical form for the profile of the foreground wedge, as a function of $\eta$, and reduces to previously-derived special cases for the wedge profile in the limit of top-hat primary beamshapes and bandpasses \citep{Trott2012}.  Had we retained the $\phi^2$ terms, they would have enforced the condition that $u \approx b \nu_0 / c$.  We therefore see that foreground contamination ought to leak to higher $\eta$ for higher values of $u$, since those are probed only by larger $b$.  Importantly, we emphasize that this foreground wedge feature is basis-independent, in that the final result makes no mention of delays.  Our choice in this paper to express visibilities and covariances using a delay basis (rather than, say, a frequency basis, i.e. spectra) is a choice that is computationally convenient (see Appendix \ref{appendix:Computational} for details), but is fundamentally an arbitrary one.  The same is true regarding our indexing of measurements by baseline.  [The baseline length $b$ that appears in  Eq. (\ref{eq:anotherWedge}) is an expression of array configuration rather than basis choice; indeed, Eq. (\ref{eq:simplifiedWind}) shows that all baseline indices are summed over].  \textit{The foreground wedge is purely a function of our instrument's design and the form of our power spectrum estimator.}

With multiple baselines,  Eq. \eqref{eq:WedgeWindow} contains cross-terms between different baselines.  Working once again at low $\eta$ and high $u$ (long baselines), one has
\begin{widetext}
\begin{eqnarray}
W_{\eta_\alpha}(\eta_\beta) \propto && \, \sum_{b_i} \frac{1}{b_i^4} \phi^4 \left( \frac{\frac{u c}{b_i} - \nu_0}{B_\textrm{band}} \right) \left( ( A_{b\parallel} \ast A_{b\parallel}) \left[ \frac{c}{b_i \theta_0} (\eta_\alpha - \eta_\beta) \right]\right)^2  +  \sum_{b_i} \sum_{b_j > b_i} \frac{2}{b_i^2 b_j^2}  \phi^2 \left( \frac{\frac{u c}{b_i} - \nu_0}{B_\textrm{band}} \right) \phi^2 \left( \frac{\frac{u c}{b_j} - \nu_0}{B_\textrm{band}} \right)\nonumber \\
&& \qquad \times \cos \left[ 2\pi \left( \frac{1}{b_i} - \frac{1}{b_j}  \right)(\eta_\alpha - \eta_\beta) u\right] ( A_{b\parallel} \ast A_{b\parallel}) \left[ \frac{c}{b_i \theta_0} (\eta_\alpha - \eta_\beta) \right] ( A_{b\parallel} \ast A_{b\parallel}) \left[ \frac{c}{b_j \theta_0} (\eta_\alpha - \eta_\beta) \right].
\end{eqnarray}

While this expression is certainly more complicated than the one we had before, the same basic picture holds: the window functions can be quite broad in the $\eta$ direction, thus allowing foregrounds to be scattered from low to medium values of $\eta$.  But in general, the resulting contamination is still enveloped by terms like $( A_{b\parallel} \ast A_{b\parallel}) \left[ \frac{c}{b \theta_0} \eta  \right]$, which limits the possible contamination at high $\eta$.

Let us now turn briefly to the behavior of the window functions as a function of $u$.  With the short baseline limit already provided by Eq. \eqref{eq:ShortBlAnalyticalWindow}, we once again focus on the long-baseline limit.  Setting $\eta = \eta_\alpha = \eta_\beta$ to isolate the $u$-dependence, Eq. \eqref{eq:LongBlSomeLimit} becomes
\begin{equation}
\label{eq:LongBlSpatialBehavior}
W_{u_\alpha}^\textrm{long bl} (u_\beta) \propto \widetilde{\gamma}^4(B_\textrm{chan} \eta)
\Bigg{|} \widetilde{A_{b\parallel}^2} \left[ \theta_0 (u_\beta - u_\alpha) \right] \sum_b \frac{c}{b\theta_0}  \phi \left( \frac{u_\alpha c / b-\nu_0}{B_\textrm{band}}\right) \phi \left( \frac{u_\beta c / b-\nu_0}{B_\textrm{band}}\right)  \Bigg{|}^2,
\end{equation}
\end{widetext}
where $\widetilde{A_{b\parallel}^2}$ denotes the Fourier transform of the square of $A_{b\parallel}$, not the square of the Fourier transform.  As expected, a baseline $b$ roughly probes a $u$-scale equal to $b \nu_0 / c$.  Additionally, the window functions peak at $u_\alpha = u_\beta$, a fact that is enforced by the appearance of the two copies of $\phi$ as a product, as well as by the presence of $ \widetilde{A_{b\parallel}^2} \left[ \theta_0 (u_\beta - u_\alpha) \right]$.  The width of the window function in the $u$ direction therefore depends on both the primary beam and the bandpass function.  The $ \widetilde{A_{b\parallel}^2}$ term has a characteristic width of $\theta_0^{-1}$, while the $\phi$ functions have a characteristic width of $B_\textrm{band} b / c$.  Now, the window function involves the product of these functions, which means that the width of its central portion will be determined mostly by the \emph{narrower} of the two contributions.  In the long baseline limit that we are working in, $\theta_0^{-1} \ll B_\textrm{band} b / c$ by definition, so the $ \widetilde{A_{b\parallel}^2}$ part is the narrower contribution and determines the width of central portion of the window function.  We thus predict that the central width is just $\theta_0^{-1}$, and does not depend on the $u$ (or equivalently, $k_\perp$) value on which the window function is centered.

However, simply characterizing the width of the central peak is insufficient for our purposes.  As emphasized throughout this paper, the large dynamic range that exists between the bright foreground emission and the dim cosmological signal means that it is important to accurately capture the weak, low-level wings of the window functions, away from the central peak.  These wings will be controlled by the broader contribution in Eq. \eqref{eq:LongBlSpatialBehavior}, namely, the product of the bandpasses.  As stated above, the bandpasses have a characteristic width $B_\textrm{band} b / c$, and since baselines of length $b$ probe spatial scales given by $u \sim  \nu_0 b / c$, our window function wings will have a width $\Delta u$ of
\begin{equation}
\label{eq:ConstantLogWidth}
\Delta u = \frac{B_\textrm{band}}{ \nu_0} u.
\end{equation}
The widths of the window function wings are therefore proportional to $u$, and grow as $u$ increases.  Equivalently, since $\Delta u \propto u$, the fractional wing width is constant, and the wings will appear to have the same width on a logarithmic $u$ (or $k_\perp$) scale.  This is intuitively unsurprising, as longer baselines probe a greater spread of spatial scale due to their greater chromaticity, and the width of this spread is proportional to the baseline length (see Figure \ref{fig:ChromaticBaselines}).  Since a given $u$ mode is mostly accessed by baselines of length $b \sim u \lambda$, one is then driven to the conclusion that $\Delta u \propto u$.

\section{A numerical model of a basic estimator}
\label{sec:basicEstNumerical}

Having made various approximations in the previous section to enable an analytic treatment of the foreground wedge, we will now discard most of these approximations in lieu of an exact numerical treatment of our basic estimator.  We will find that the basic picture that we presented above remains unchanged.

\subsection{Instrument and foreground model}
The model instrument that we consider in this paper is intended to reflect a typical design for an interferometer optimized for $21\,\textrm{cm}$ power spectrum (rather than one that is intended to function as a general-purpose low-frequency radio observatory).  Maximizing power spectrum sensitivity requires antennas to be placed in a way that yields a large number of short, identical baselines \cite{Parsons2012a,Pober2014}.  With this in mind, we perform our computations for a square, $20$ by $20$ array of antennas, with a $14\,\textrm{m}$ spacing between adjacent antennas.  Each antenna is assumed to have a Gaussian primary beam, with a full-width-half-max (FWHM) of $40.5^\circ$ that is approximated as frequency-independent in the $B_\textrm{band} = 8\,\textrm{MHz}$ band that we consider.  The frequency width of each individual spectral channel is set at $B_\textrm{chan} = 50\,\textrm{kHz}$.  With no loss of qualitative generality, we consider only observations centered around $\nu_0 = 150\,\textrm{MHz}$.  The formalism presented in this paper applies to all redshifts accessible to a $21\,\textrm{cm}$ interferometer, and none of the ``lessons learned" in our analysis are substantially changed by examining a different redshift.

For computational simplicity, we take the tapering function $\phi [ (\nu - \nu_0) / B_\textrm{band} ]$ of our delay transform [Eq. \eqref{eq:DelayDef}] to be Gaussian, even though previous studies in the literature have argued for more desirable choices such as Blackman-Harris function or a Blackman-Nuttal function \cite{Thyagarajan2013,Parsons2013}.  Using a Gaussian allows us to compute analytically compute the $\nu$ integral in our measurement equation [Eq. \eqref{eq:vIntegratedGenVtilde}], giving:
\begin{widetext}
\begin{eqnarray}
\widetilde{V} (\mathbf{b}, \tau) = 2 \pi \theta_0^2 B_\textrm{band} \int d^2 \mathbf{u} d\eta \, \widetilde{I} (\mathbf{u}, \eta)  \exp \left[ - \frac{2 \pi^2 \theta_0^2 \left(u - \nu_0 b / c\right)^2}{1+ \left( 2 \pi \theta_0 b B_\textrm{band}/c\right)^2}  \right] \exp \left[ - \frac{2 \pi^2 B_\textrm{band}^2 \left( \eta- \tau \right)^2}{1+ \left( 2 \pi \theta_0 b B_\textrm{band}/c\right)^2}  \right] \nonumber \\
\times \exp \left[ i 2 \pi B_\textrm{band} (\eta - \tau) \frac{4 \pi^2 \theta_0^2 \left(u - \nu_0 b / c\right)\left( b B_\textrm{band} / c\right)}{1+ \left( 2 \pi \theta_0 b B_\textrm{band}/c\right)^2}\right] \frac{ \widetilde{A}_{b\perp} (\theta_0 v) \widetilde{\gamma} (B_\textrm{chan} \eta) }{\sqrt{1+ \left( 2 \pi \theta_0 b B_\textrm{band}/c \right)^2}} e^{i 2 \pi \nu_0 (\eta - \tau)},
\end{eqnarray}
where the characteristic scale of the beam $\theta_0$ is given by the standard deviation of our Gaussian beam, which in our case is $\theta_0 = \textrm{FWHM} / \sqrt{8 \ln 2} = 17.2^\circ$.  Intuitively, one sees that each delay mode of each baseline probes a reasonably localized region in $uv\eta$ space.  One also sees that there exists a complex exponential term that mixes spatial and spectral information, which is to be expected given the chromatic nature of an interferometer's synthesized beam.

Following this, we form the covariance matrix $\mathbf{C}_{ij} \equiv \langle \widetilde{V} (\mathbf{b}_i, \tau_i) \widetilde{V}^* (\mathbf{b}_j, \tau_j) \rangle$ under the same coherency approximation as the one we employed in the previous section: two baselines may have any amount of overlap on the $uv$ plane in the direction parallel to their baseline vector, but are either completely non-overlapping or perfectly overlapping in the direction perpendicular to the baseline vector.  This allows the integral over $v$ (defined to be the direction on the $uv$ plane perpendicular to a pair of correlated baselines) to be evaluated analytically.  The sky signal portion of our covariance matrix is then
\begin{eqnarray}
\label{eq:GaussianCovar}
\mathbf{S}_{ij} = && 4 \pi^\frac{5}{4} B_\textrm{band}^2 \theta_0^4 \, e^{i 2 \pi \nu_0 (\tau_j - \tau_i)} \int du d\eta \frac{P(u,\eta) }{\sqrt{1+ \alpha_i^2} \sqrt{1+ \alpha_j^2}} \,\widetilde{\gamma}^2 \!( B_\textrm{chan} \eta ) \exp \left[ - 2 \pi^2 \theta_0^2 \left( \frac{\left(u - \nu_0 b_i / c\right)^2}{1+ \alpha_i^2} + \frac{\left(u - \nu_0 b_j / c\right)^2}{1+ \alpha_j^2} \right)\right] \nonumber \\
&& \times \exp \left[ - 2 \pi^2 B_\textrm{band}^2 \left( \frac{\left( \eta - \tau_i \right)^2}{1+ \alpha_i^2} + \frac{\left( \eta - \tau_j \right)^2}{1+ \alpha_j^2} \right)\right] \exp \left[ i 4 \pi^2 \left( \alpha_i  \frac{\left( \eta - \tau_i \right)  \left(u - \nu_0 b_i / c\right) }{1+ \alpha_i^2} - \alpha_j  \frac{\left( \eta - \tau_j \right)  \left(u - \nu_0 b_j / c\right)}{1+ \alpha_j^2} \right)\right], \qquad 
\end{eqnarray}
where $\alpha_i \equiv 2\pi \theta_0 B_\textrm{band} b_i / c$ and similarly for $\alpha_j$.  We take the frequency channel response $\gamma$ to be a Gaussian, which makes its Fourier transform $\tilde{\gamma}$ also a Gaussian.  Computing $\mathbf{C}_{,\alpha}$ is very similar to computing $\mathbf{S}$.  Since instrumental noise is random and does not depend on the power spectrum, we have $\mathbf{C}_{,\alpha} = \mathbf{S}_{,\alpha}$.  To find $\mathbf{C}_{,\alpha}$, then, we simply need to evaluate the integrals in Eq. \eqref{eq:GaussianCovar}, but with $u$ and $\eta$ integration limits chosen to match to the band in question, rather than being $-\infty$ and $+\infty$.
\end{widetext}
Having described our instrument and how it manifests itself in the covariance of our measurements, the final ingredient that we require for our numerical calculations is a model for the total power spectrum $P(u,\eta)$.  We model the total power spectrum as the sum of the cosmological power spectrum and a foreground power spectrum.  For the cosmological power spectrum, we use the spherically symmetric power spectrum provided in Ref. \cite{Barkana2009} and assume statistical isotropy to compute the cylindrical power spectrum needed for our covariance.  As for the foregrounds, we consider a relatively simple two-component power spectrum model $P_\textrm{fg}$:
\begin{equation}
\label{eq:PowerSpectrumForegrounds}
P_\textrm{fg} (u, \eta) = A \left( C_{\ell = 2 \pi u}^\textrm{diff} e^{- \nu_c^\textrm{diff} |\eta|} + C_{\ell = 2 \pi u}^\textrm{ps} e^{- \nu_c^\textrm{ps} |\eta|} \right),
\end{equation}
where $A$ is an overall normalization, $C_\ell^\textrm{diff}$ is the angular power spectrum of the diffuse Galactic emission, and $\nu_c^\textrm{diff}$ is its frequency coherence length.  The corresponding quantities for point sources are given by $C_{\ell = 2 \pi u}^\textrm{ps}$ and $ \nu_c^\textrm{ps}$.  The $\eta$ dependence of this parametric form is motivated by the mathematical results of Ref. \cite{Liu2012}.  In that work, empirically-motivated models of foreground spectra were put through an eigenmode analysis, and it was found that the resulting set of eigenmodes were essentially Fourier modes in frequency (i.e. $\eta$ modes).  The eigenvalue spectra were well-fit by a linear exponential in $\eta$, with a coherence frequency of $64.8\,\textrm{MHz}$.  For simplicity, we adopt this value for both  $ \nu_c^\textrm{ps}$ and  $ \nu_c^\textrm{diff}$.

To model the angular structure of the diffuse Galactic emission, we use the Global Sky Model (GSM) software \cite{deOliveiraCosta2008} to generate a model of the sky at $150\,\textrm{MHz}$.  We then compute the angular power spectrum of this model, which we find to be well-fit by
\begin{equation}
C_\ell^\textrm{diff} \propto
\begin{cases}
\exp \left( a_1 \ell + a_2 \ell^2\right) &\, \textrm{for } \, \ell \le 8 \\
b_1 \ell^{b_2} &\, \textrm{for }\,\ell > 8,
\end{cases}
\end{equation}
with $a_1 = -1.450$, $a_2 = 0.1003$, $b_1 = 0.7666$, and $b_2 = -2.365$.  For notational simplicity, we have omitted the overall normalization of $C_\ell^\textrm{diff}$, as it can be absorbed into $A$.  We note that while only the high $\ell$ (power-law) portion of the angular power spectrum is typically modeled in foreground studies, it is crucial to include the low $\ell$ behavior as well.  To see this, note that  Eq. \eqref{eq:GaussianCovar} predicts a substantial overlap between the response of the shortest baselines and the $u=0$ mode of the power spectrum. This is simply reflecting the fact that the sky is not infinite in extent.  Thus, even though the auto-correlation/zero-spacing baseline products are typically discarded from interferometric data, the instrument may still be sensitive to the zero mode of the sky.

For the point source contribution to foregrounds, we neglect clustering for simplicity, and therefore take $C_{\ell = 2 \pi u}^\textrm{ps}$ to be a constant.  We fix this constant by assuming that at $\ell = 1000$, the amplitude of the point source angular power spectrum is roughly a factor of 10 smaller than that of the diffuse emission \cite{Santos2005}.  We therefore set $C_{\ell = 2 \pi u}^\textrm{ps} = 0.1 \,C_{\ell=1000}^\textrm{diff}$.  Using a simple, $\ell$-independent angular power spectrum for point sources is not a required approximation for our formalism, and this assumption can be easily relaxed.  In principle, including clustering would boost the point source power at low $\ell$ modes \cite{DiMatteo2002}.  In practice, however, the low $\ell$ regime is dominated by the diffuse Galactic emission anyway, and we do not expect that an inclusion of clustering would qualitatively impact our numerical results.

We fix the overall normalization $A$ of our foregrounds by considering the zero-mode of the power spectrum.  We require that
\begin{equation}
P(u=0,\eta=0) = B_\textrm{band} \theta_0^2 \,\overline{I}_{GSM}^2,
\end{equation}
where $\overline{I}_{GSM} = 433\,\textrm{K}$ is the mean temperature of our GSM foreground template.

Finally, we must add the noise contribution to our covariance.  The computation of a noise covariance matrix $\mathbf{N}$ is rather subtle, given the assumptions that we have made above regarding baselines that overlap in directions perpendicular to their baseline vectors.  We first sort the baselines of our array by baseline length into 54 equally-spaced bins.  If only one baseline fell into each bin, the noise variance assigned to each bin would be \cite{Parsons2012a}
\begin{equation}
\label{eq:NoiseCovarSingelBl}
\mathbf{N}_{ii} \Big{|}_\textrm{single bl} = \frac{\Omega_{\textrm{pp}}}{2t}  B_\textrm{band} T_\textrm{sys}^2,
\end{equation}
where $t$ is the integration time (taken to be $520\,\textrm{hrs}$), $T_\textrm{sys}$ is the system temperature, and
\begin{equation}
\label{eq:Omegapp}
\Omega_\textrm{pp} \equiv  \int A^2(\boldsymbol\theta  / \theta_0) d^2 \theta.
\end{equation}
Note that our expression for the noise variance differs from equations that are commonly seen in the literature in two ways.  First, our variance is proportional to $B_\textrm{band}$.  This is simply due to the fact that we are working in a delay basis rather than a frequency basis.  In addition, the beam area used here is the integrated \emph{square} of the beam profile (rather than just the integral of the beam profile itself), which was shown in Ref. \cite{Parsons2013} to be the correct beam area to use for this calculation.  With our Gaussian primary beam, $\Omega_\textrm{pp} = \pi \theta_0^2$.  We assume a sky-noise dominated instrument and set $T_\textrm{sys} = \overline{I}_{GSM}$.

We now adjust for the fact that each baseline length bin contains more than just a single baseline.  If there are $n(b_i)$ baselines within a particular bin, we simply divide the noise variance by $n(b_i)$, assuming that a combination of instantaneous redundancy and rotation synthesis allow a large fraction of the baselines to be combined coherently, prior to forming a power spectrum.  (In practice, this is a somewhat optimistic assumption, given that baselines that are dissimilar in orientation can only be combined statistically, and not coherently).  Further assuming that the noise covariance matrix is diagonal, our final form for $\mathbf{N}$ is
\begin{equation}
\mathbf{N}_{ij} =\frac{1}{n(b_i)} \frac{\Omega_{\textrm{pp}}}{2t}  B_\textrm{band} T_\textrm{sys}^2 \delta_{ij},
\end{equation}
where the Kronecker delta function ensures that instrumental noise is uncorrelated both between baselines and between delay bins.

Admittedly, the instrument and noise models presented in this section are quite crude and make use of a large number of simplifying assumptions.  The assumptions regarding instrumental noise and rotation synthesis, in particular, are quite optimistic.  However, the numerical computations that we perform in this paper are not designed to be definitive sensitivity calculations.  Rather, the goal is to use our rough model---which captures the essential features of large-amplitude smooth-spectrum foregrounds and lower-amplitude broadband instrumental noise---to gain some statistically rigorous intuition for the EoR window.  For further technical details regarding our exact implementation (e.g. for information about bin sizes), we refer the reader to Appendix \ref{appendix:Computational}.

\subsection{Window functions and foreground bias}

\begin{figure}[t] 
	\centering 
	\includegraphics[width=.49\textwidth]{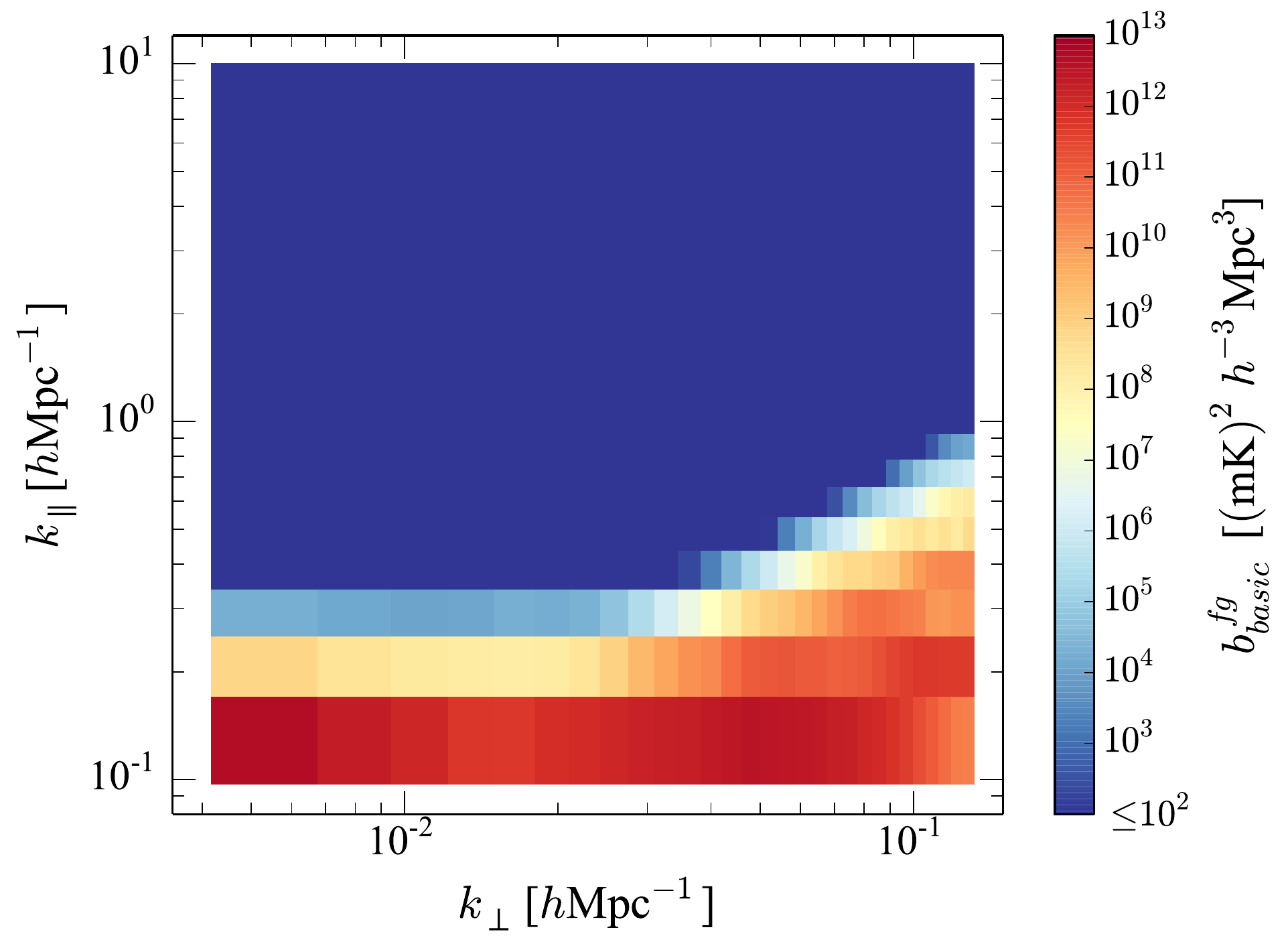}
	\caption{Expected foreground bias [Eq. \eqref{eq:fgBias}] for the basic estimator defined in Section \ref{sec:basicEstAnalytic}, where $\mathbf{E}_\alpha \propto \mathbf{N}^{-1} \mathbf{C}_{,\alpha} \mathbf{N}^{-1}$.  For this basic estimator, the chromatic nature of the instrument results in foreground contamination in the form of a characteristic wedge at high $k_\perp$.}
	\label{fig:basicEstBias}
\end{figure} 
\begin{figure*}[!ht] 
	\centering 
	\includegraphics[width=1\textwidth]{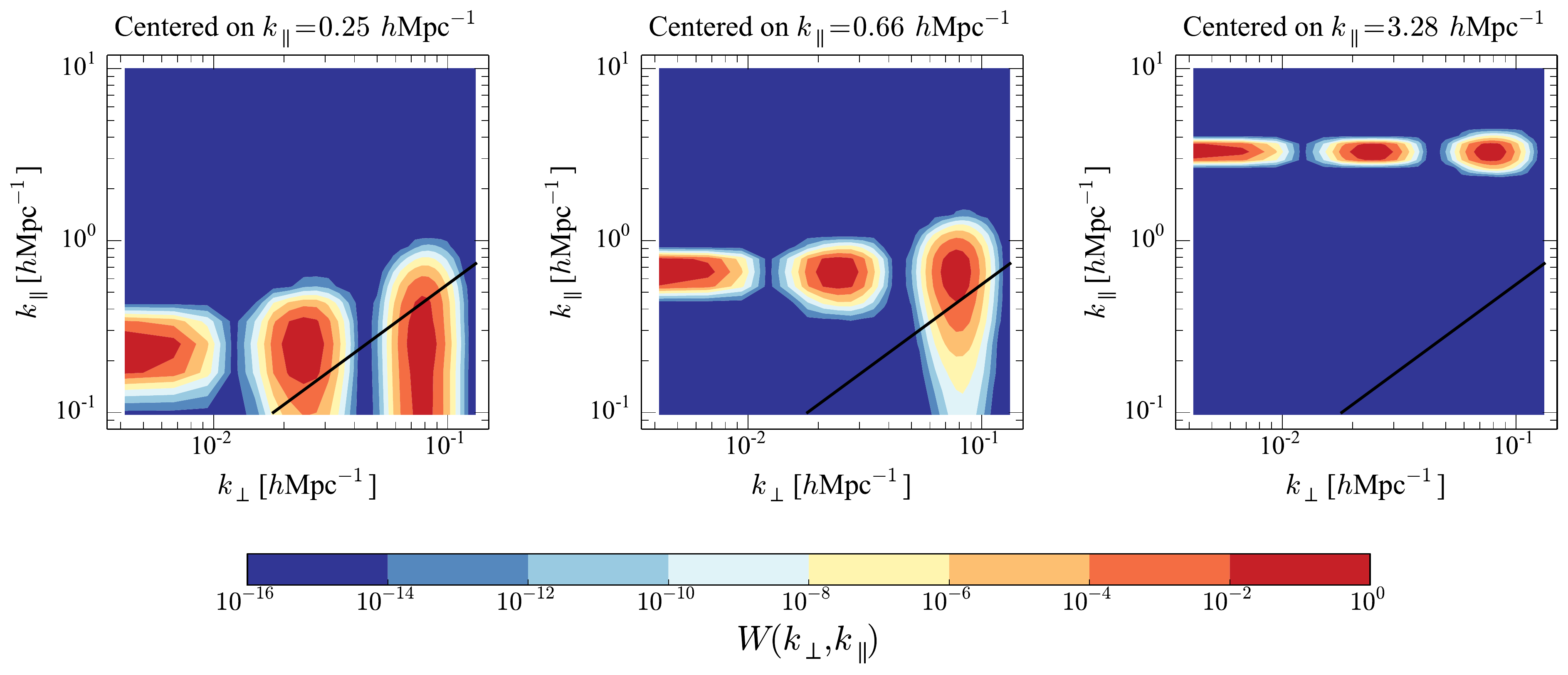}
	\caption{Sample window functions on the $k_\perp k_\parallel$ plane for the basic estimator defined by $\mathbf{E}_\alpha \propto \mathbf{N}^{-1} \mathbf{C}_{,\alpha} \mathbf{N}^{-1}$.  The leftmost plot shows a set of window functions centered at $k_\parallel = 0.25\,h$Mpc$^{-1}$; the middle plot shows a set centered at $k_\parallel = 0.66\,h$Mpc$^{-1}$; the rightmost plot at $k_\parallel = 3.28\,h$Mpc$^{-1}$.  Within each plot, the window functions are centered, from left to right, at $k_\perp = 0.0042\,h$Mpc$^{-1}$, $k_\perp = 0.024\,h$Mpc$^{-1}$, $k_\perp = 0.078\,h$Mpc$^{-1}$.  The black solid line in each plot indicates the rough extent of the foreground wedge [Eq. \eqref{eq:CosmologicalCoordsWedge} with $\theta_0 = \pi / 2$].  For window functions centered at high $k_\perp$, there is substantial elongation in the $k_\parallel$ direction, causing higher $k_\parallel$ modes to pick up foreground power, resulting in a foreground-contaminated wedge.}
	\label{fig:basicEstWindows}
\end{figure*}

We now examine the statistical properties of our basic power spectrum estimator.  We will find that with our basic estimator, the foregrounds appear in a wedge in $k_\perp k_\parallel$ space, consistent with findings in the previous literature.

In Figure \ref{fig:basicEstBias}, we show the expected foreground bias from  Eq. \eqref{eq:fgBias} in terms of the cylindrical Fourier coordinates\footnote{In the mathematical formalism that we have developed so far, we have preferred to use $u$ and $\eta$ as our Fourier coordinates perpendicular and parallel to the line-of-sight, respectively.  However, to better relate our results to theoretical predictions, we convert to using the more conventional Fourier coordinates $k_\perp$ and $k_\parallel$ when displaying our results.  For details about this conversion and our Fourier conventions, please see Appendix \ref{appendix:Fourier}.} $k_\perp$ and $k_\parallel$.  This shows the expected level of additive foreground bias in an estimate of the power spectrum.  This foreground bias is the quantity that is most directly comparable to previous studies where a single set of simulated foregrounds are propagated through a power spectrum estimation pipeline.  We see that our results are consistent with such studies as well as the analytic arguments of the previous section: the foregrounds are mostly sequestered to low $k_\parallel$ modes at low values of $k_\perp$, rising to higher $k_\parallel$ modes at high $k_\perp$ in a wedge-like pattern.  As predicted by  Eq. \eqref{eq:CosmologicalCoordsWedge}, the edge of the wedge is defined by a line with unit logarithmic slope.  Beyond this edge, the contamination drops off sharply to give a clean EoR window.

Importantly, we emphasize that the characteristic shapes seen here are not tied to the specifics of our foreground model, beyond the fact that foregrounds are spectrally smooth.  Instead, the foreground wedge is tied to the chromatic nature of our instrument and the properties of our basic estimator.  While this may be difficult to establish definitively with a simulation, the covariant formalism that we make use of in this paper allows the foreground models to be easily disentangled from the way they interact with the instrument and the power spectrum estimation pipeline.  For example, in Figure \ref{fig:basicEstWindows}, we show some examples of window functions on the $k_\perp k_\parallel$ plane.  From  Eq. \eqref{eq:WindGeneral}, we know that these functions depend only on our choice of estimator (through $\mathbf{E}_\alpha$) and the instrument's response (through $\mathbf{C}_{,\beta}$).  Thus, any signatures of the wedge that are independent of the foreground model should be apparent in the window functions.

Consider first the leftmost plot from Figure \ref{fig:basicEstWindows}, which shows three window functions that are all centered on the $k_\parallel$ value of $0.25\,h$Mpc$^{-1}$, but different $k_\perp$ values.  As one moves to higher $k_\perp$, the window functions become increasingly elongated\footnote{This elongation is real and not just an artifiact of our logarithmic $k_\perp k_\parallel$ axes.  We know this because the set of three windows shown in each plot of Figure \ref{fig:basicEstWindows} are chosen to be centered on the same $k_\parallel$.} in the $k_\parallel$ direction.  With long tails to low $k_\parallel$ (where foreground emission naturally resides), this implies a leakage of smooth foregrounds from low to high $k_\parallel$.  Since this effect is most pronounced at high $k_\perp$, the result is precisely a wedge-like structure.  To guide the eye, the black lines  on each plot show the edge of the wedge as predicted\footnote{Our use of $\theta_0 = \pi/2$ to define the edge of the wedge is somewhat arbitrary, given that there is nothing special about $\theta_0 = \pi / 2$ in our flat-sky approximation.  However, as emphasized in Ref. \cite{Parsons2012b}, in a proper curved-sky treatment it is the natural scale to consider, since the primary beam of an instrument must vanish at the horizon.} by  Eq. \eqref{eq:CosmologicalCoordsWedge} with $\theta_0 = \pi / 2$.

Window functions centered at higher $k_\parallel$ values (central and rightmost plots in Figure \ref{fig:basicEstWindows}) also develop elongations as one moves from low to high $k_\perp$, although the effect is visually subtle due to our logarithmic plotting.  These elongations are slightly less important for smooth foregrounds, as even the elongated tails are not quite long enough to reach the lowest $k_\parallel$ for window functions that are centered at high $k_\parallel$.  However, such effects may be important if foregrounds turn out to contain unsmooth (high $k_\parallel$) components (see Section \ref{sec:Unsmooth} for a brief discussion of this).

Except for at the lowest $k_\perp$ values, the width of the window functions in the $k_\perp$ direction also appears to increase with increasing $k_\perp$.  Plotted on the logarithmic axes of Figure \ref{fig:basicEstWindows}, it is visually clear that at intermediate to high $k_\perp$ the window functions have a roughly constant \emph{logarithmic} width, confirming the proportional increase in width with $k_\perp$ that we predicted in Eq. \eqref{eq:ConstantLogWidth}.

Our analytic, covariant treatment of power spectrum statistics allows us to compute window functions to the high dynamic range shown in Figure \ref{fig:basicEstWindows}.  This is crucial given that foregrounds are expected to be $\sim 10^{10}$ times brighter in power (i.e., in temperature-squared units) compared to the cosmological signal.  It is therefore essential to capture the low-level tails of window functions.  Conveniently, once the windows have been computed, the foreground bias for a different foreground power spectrum can be easily determined using  Eq. \eqref{eq:WindToBias}.

\subsection{Error bars and error covariance}

\begin{figure}[t] 
	\centering 
	\includegraphics[width=.49\textwidth]{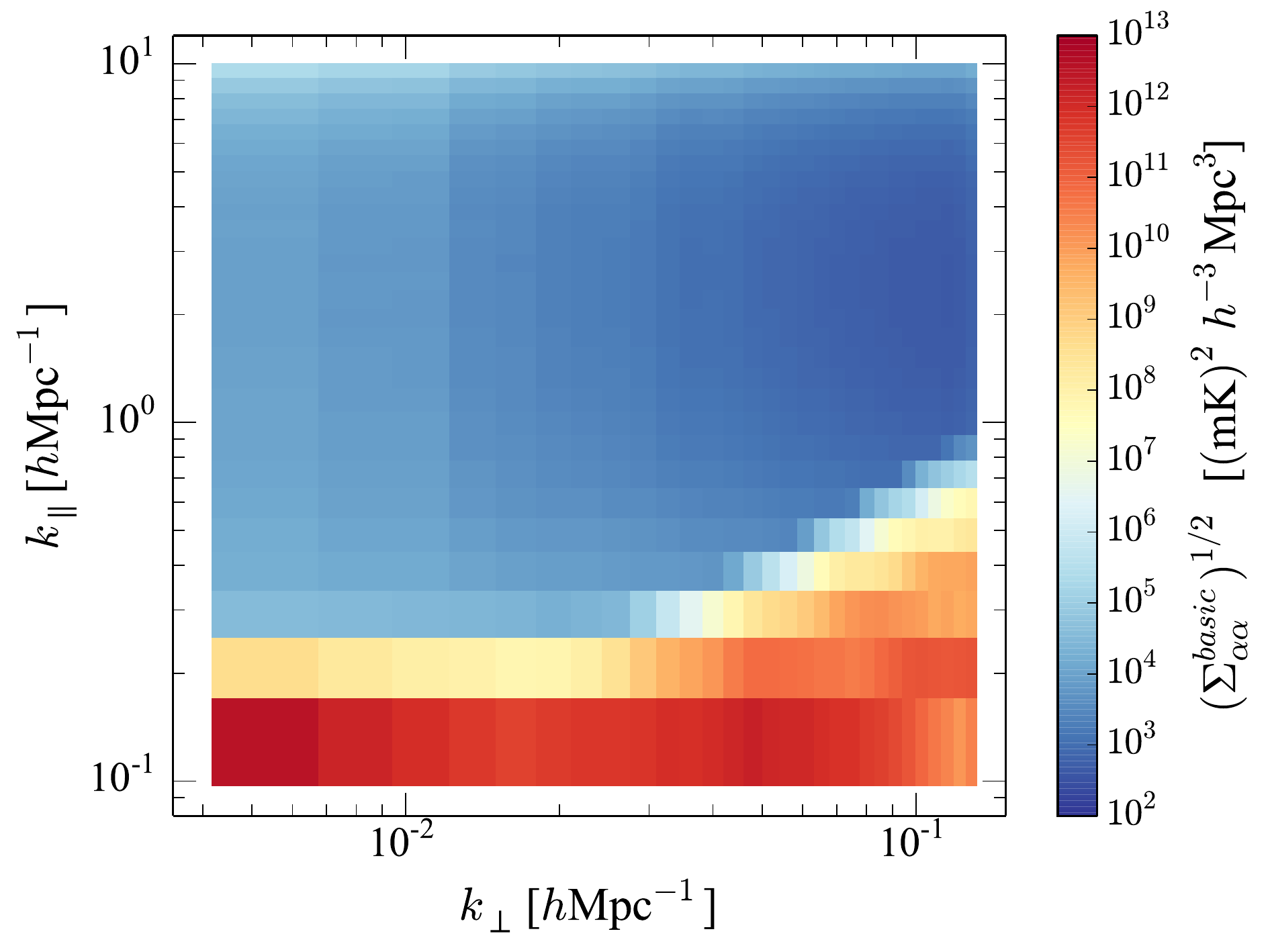}
	\caption{Expected foreground error bars [given by $\left(\boldsymbol \Sigma_{\alpha\alpha}\right)^{\frac{1}{2}}$ for the $\alpha$th bandpower or $k_\perp k_\parallel$ cell] for the basic estimator defined by $\mathbf{E}_\alpha \propto \mathbf{N}^{-1} \mathbf{C}_{,\alpha} \mathbf{N}^{-1}$.  The foreground wedge also shows up in the error bars of a power spectrum measurement.}
	\label{fig:basicEstErrorBars}
\end{figure}

In addition to the foreground bias, we may quantify the error covariance in estimates of the power spectrum.  In Figure \ref{fig:basicEstErrorBars}, we show the square-root of the diagonal elements of the error covariance matrix [Eq. \eqref{eq:matrixErrorCovar}], i.e. the power spectrum error bars.  These error bars capture more than just thermal noise errors, and include contributions from the foreground covariance.  Indeed, one sees that the foreground wedge appears not just in the bias that we discussed previously, but also in the form of increased error bars.  Outside the wedge, the error bars are dominated by thermal noise, and are quite low, in what constitutes the EoR window.  Note that the thermal noise contribution did not appear in the bias, since we have assumed that cross-correlations have eliminated the noise bias.  Towards the smallest $k_\perp$, the errors increase by a small amount due to cosmic variance.  This effect is typically negligible unless $k_\parallel$ is also small, but does play a small role since our model has a rather low noise level.  (Again, we emphasize that our goal is not to perform a definitive sensitivity calculation).  At the highest $k_\parallel$, the errors also rise slightly.  This is due to the finite spectral resolution of our instrument, which is self-consistently included in our error bars via the spectral channel profile $\gamma$ in our formalism.

\begin{figure*}[!ht] 
	\centering 
	\includegraphics[width=1\textwidth]{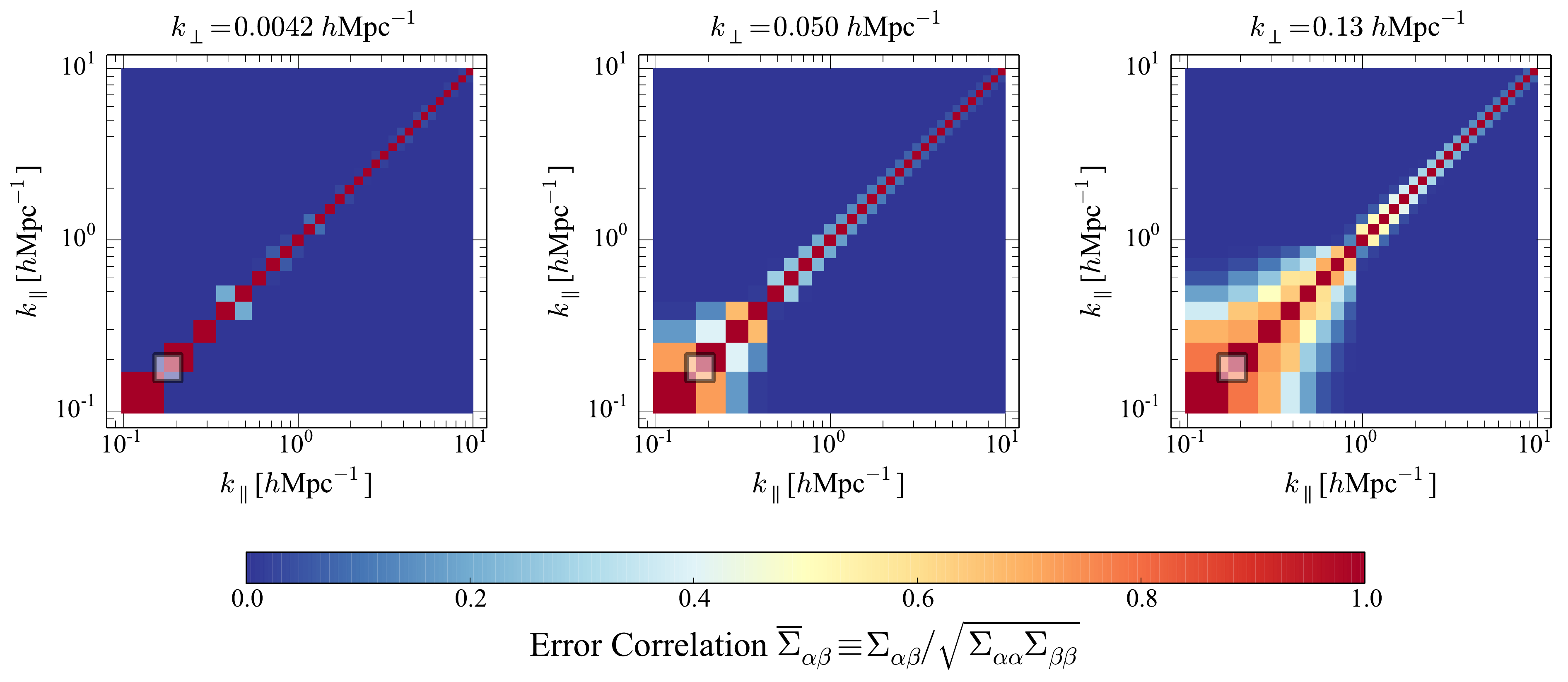}
	\caption{Sections of the measurement error correlation matrix [Eq. \eqref{eq:ErrorCorr}] for our basic $\mathbf{E}_\alpha \propto \mathbf{N}^{-1} \mathbf{C}_{,\alpha} \mathbf{N}^{-1}$ estimator.  Each plot shows the measurement error correlation between all $k_\parallel$ bins for fixed $k_\perp$ (from left to right, $k_\perp = 0.0042\,h$Mpc$^{-1}$, $0.050\,h$Mpc$^{-1}$, and $0.13\,h$Mpc$^{-1}$).  It is often assumed in the literature that errors in two $k_\parallel$ cells are uncorrelated if the cells are more than $\Delta k_\parallel \sim 2\pi \frac{ H_0 E(z)}{c (1+z)^2} \frac{1}{B_\textrm{band}}$ apart.  Overlaid near the bottom left corner of each plot is a semi-transparent square of size $\Delta k_\parallel \times \Delta k_\parallel$.  Comparing the sizes of these squares to the width of the off-diagonal correlations, one sees that such an assumption holds only at low $k_\perp$.  At high $k_\perp$ the errors are highly correlated, reducing the number of independently-measurable modes.}
	\label{fig:basicEstCovar}
\end{figure*}

Beyond just the error bars, our formalism also delivers the off-diagonal elements of the error covariance $\boldsymbol \Sigma$, which encode the error correlations between different $k_\perp k_\parallel$ cells.  As emphasized in Ref. \cite{Dillon2014}, these correlations need to be quantified if one wishes to accurately propagate errors from the cylindrical power spectrum $P(k_\perp, k_\parallel)$ to the spherical power spectrum $P(k)$.  To capture this information, we consider the correlation matrix, defined as
\begin{equation}
\label{eq:ErrorCorr}
\overline{\boldsymbol \Sigma}_{\alpha \beta} \equiv \frac{\boldsymbol \Sigma_{\alpha \beta}}{\sqrt{\boldsymbol \Sigma_{\alpha \alpha} \boldsymbol \Sigma_{\beta \beta}}},
\end{equation}
which is essentially a whitened version of the error covariance.  Examining $\overline{\boldsymbol \Sigma}$ instead of $\boldsymbol \Sigma$ allows the correlations rather than the larger errors within the wedge to be the dominant feature.  To further aid visualization, we focus on just small portions of the correlation matrix.  Whereas the full correlation matrix would relate all $k_\perp k_\parallel$ coordinates to all other such coordinates, in Figure \ref{fig:basicEstCovar} we fix $k_\perp$ at three separate values, and consider the correlations between different $k_\parallel$ coordinates.

Immediately obvious is the fact that the error correlations form qualitatively different structures at different values of $k_\perp$.  To calibrate our expectations, we include in each plot a semi-transparent square of size $\Delta k_\parallel \times \Delta k_\parallel$, where
\begin{equation}
\label{eq:kparaCorrLen}
\Delta k_\parallel \sim 2\pi \frac{ H_0 E(z)}{c (1+z)^2} \frac{1}{B_\textrm{band}}.
\end{equation}
This is the scale over which errors are expected to be correlated in the $k_\parallel$ direction.  It is derived by making the assumption that for a survey with bandwidth $B_\textrm{band}$, the error correlation scale in $\eta$ should be roughly $1/ B_\textrm{band}$, and expressing this scale in cosmological Fourier coordinates.  In the leftmost plot of Figure \ref{fig:basicEstCovar}, where $k_\perp$ is fixed at the low value of $0.0042\,h$Mpc$^{-1}$, we see that our rough expectations are correct.  Having chosen our Fourier cell sizes with the survey volume in mind (see Appendix \ref{appendix:Computational} for details), the cells are seen to be essentially uncorrelated.

\begin{figure}[t] 
	\centering 
	\includegraphics[width=.49\textwidth]{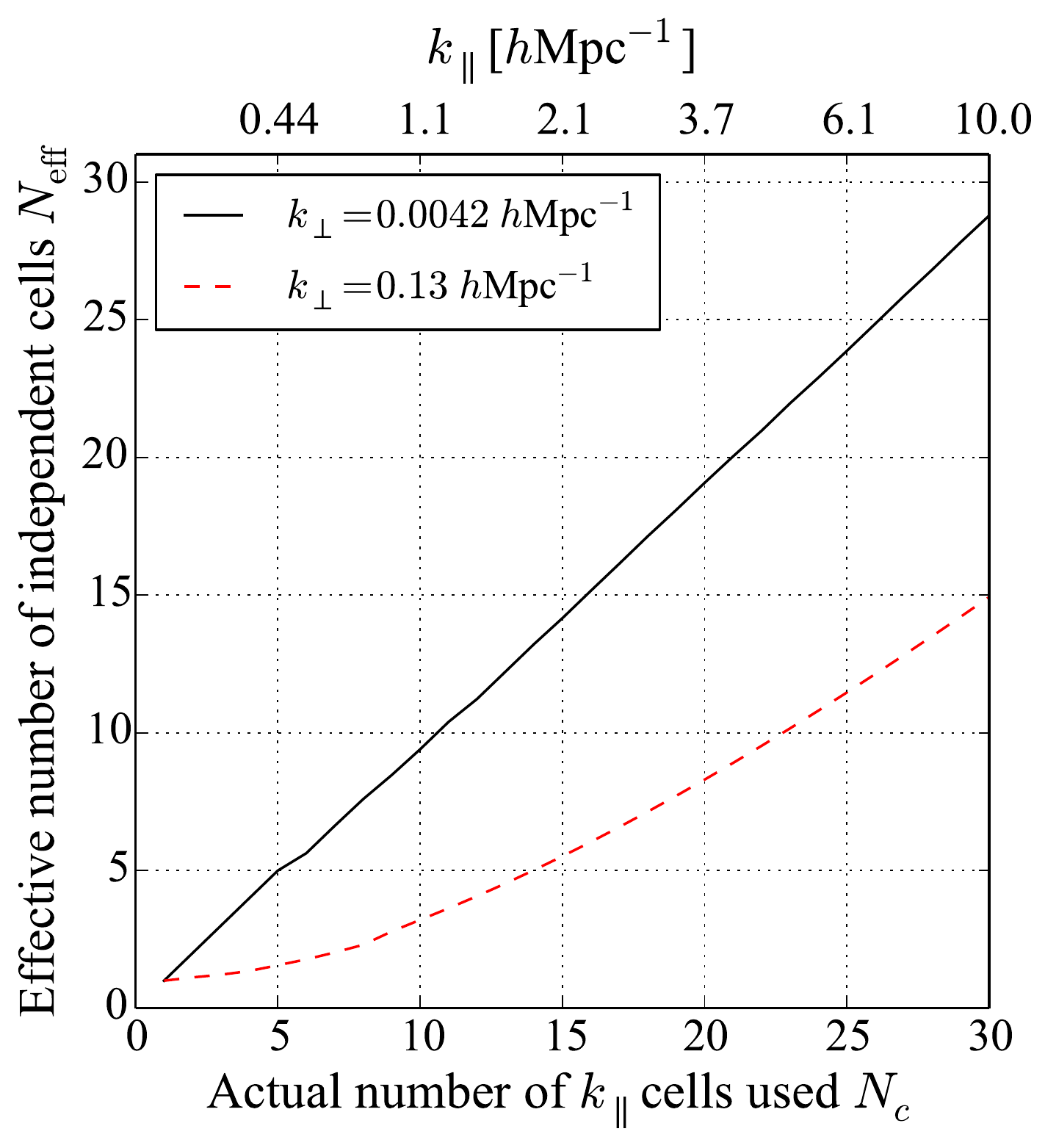}
	\caption{Effective number of independent cells $N_\textrm{eff}$ as a function of $N_c$, the number of Fourier cells included in an averaging of $k_\parallel$ modes (going from low to high $k_\parallel$).  The approximately linear relationship seen for the low $k_\perp$ curve is indicative of uncorrelated modes.  On the other hand, the high $k_\perp$ curve approaches linearity only when high $k_\parallel$ modes dominate the average.  Its initially slow increase at low $k_\parallel$ shows that the modes are highly correlated.  Thus, for power spectrum sensitivity calculations to be reliable, one must take error correlations into account or risk overestimating an instrument's sensitivity.}
	\label{fig:Neff}
\end{figure}

This simple picture breaks down, however, as we move towards higher $k_\perp$, where the effects of the foreground wedge are more pronounced.  Figure \ref{fig:basicEstCovar} shows that increasing $k_\perp$ (which essentially means moving deeper and deeper into the wedge) causes different $k_\parallel$ cells to become increasingly correlated.  To understand why error correlations are to be expected, consider what it would take for the errors to be uncorrelated in $k_\parallel$.  Suppose one had an achromatic instrument with noise properties that were uncorrelated and uniform (i.e. white) between all frequency channels.  Moving into $k_\parallel$ space by way of a Fourier transform does not induce any error correlations, because uncorrelated white noise has the same statistical properties in all bases.  On the other hand, for our measurement we have an inherently chromatic instrument, which makes our noise chromatic.  Fourier transforming non-white noise will result in noise correlations, even if the noise was uncorrelated to begin with.  Additionally, interferometers are chromatic in a very specific way, with longer baselines more chromatic (as illustrated in Figure \ref{fig:ChromaticBaselines}).  Since higher $k_\perp$ are probed by longer baselines, error correlations should increase with $k_\perp$, as seen in Figure \ref{fig:basicEstCovar}.

Viewed together, Figures \ref{fig:basicEstErrorBars} and \ref{fig:basicEstCovar}  suggest that the characteristic $k_\parallel$ correlation scale coincides roughly with the $k_\parallel$ extent of the wedge.  This suggests that the instrumental effects that caused the wedge have also decreased the number of independent measurements, thus decreasing overall signal-to-noise.  (Note, however, that the correlated errors persist even at high $k_\parallel$.  Though somewhat challenging to see with our logarithmic axes and hybrid binning, one sees that even there the off-diagonal $k_\parallel$ correlations are greater for large $k_\perp$ modes).  That there is a rough matching of scales between the wedge and our error correlation scale is not entirely surprising (though not a given in the \emph{a priori} sense\footnote{While window functions and error covariances are related to one another, they are conceptually separate entities, and window function widths do not in general coincide perfectly with error correlation lengths.  In Paper II, for example, we will see an example of an estimator with completely uncorrelated errors, but whose window functions continue to have a non-zero width.  It is also possible to write down estimators that artificially force the window functions to be delta functions, with the corresponding errors becoming \emph{anti}-correlated \cite{Dillon2014,Shaw2014b}.}), since that window functions and covariances are closely related to one another.  Recall that our window functions exhibited long elongations in $k_\parallel$ as we moved towards higher $k_\perp$.  With extremely broad window functions, our instrument essentially smoothed over a large number of modes, and it is unsurprising that the errors in nearby bins ended up being positively correlated.

We can quantify the error correlations in more detail by defining an effective number of independent cells $N_\textrm{eff}$, where
\begin{equation}
N_\textrm{eff} \equiv \frac{N_c^2}{\sum_{\alpha,\beta}^{N_c} \overline{\boldsymbol \Sigma}_{\alpha \beta}},
\end{equation}
with $N_c$ being the number of Fourier cells that enter into a simple unweighted averaging of Fourier modes.  By construction, if all the modes that one averages over are independent, $N_\textrm{eff}$ equals $N_c$, whereas $N_\textrm{eff} = 1$ if the modes are perfectly correlated.  In Figure \ref{fig:Neff}, we consider the effect of averaging Fourier modes along the $k_\parallel$ axis and show $N_\textrm{eff}$ as a function of $N_c$, with $N_c$ increasing from unity (when only the lowest $k_\parallel$ mode is included in the average) to $N_c = 30$ (the total number of $k_\parallel$ cells in our computation).  We do this for two different constant $k_\perp$ slices on the Fourier plane, $k_\perp = 0.0042\,h$Mpc$^{-1}$ and $0.13\,h$Mpc$^{-1}$ (corresponding to the leftmost and rightmost plots of Figure \ref{fig:basicEstCovar}, respectively).  For low $k_\perp$, one sees the linear relationship $N_\textrm{eff} \approx N_c$ regardless of how many $k_\parallel$ cells are included in the averaging.  For high $k_\perp$, however, $N_\textrm{eff}$ increases only very slowly at first, as one averages together the highly-correlated modes within the wedge.  The increase in $N_\textrm{eff}$ is linear only at higher $k_\parallel$, where our hybrid binning becomes logarithmic, and each cell encompasses a greater extent in $k_\parallel$, eventually exceeding the error correlation length.  In our formalism, the correlated errors within each of these larger cells have already been self-consistently averaged over, giving independent cells that contribute linearly to $N_\textrm{eff}$.  Regardless of the specific of one's power spectrum estimation formalism, it is crucial to take into account error correlations if averaging together bins that are narrower than the error correlation length, or if the bins are wider than this, to ensure that the implicit averaging performed within the bin is done correctly [rather than relying on possibly incorrect \emph{a priori} assumptions such as that implied by Eq. \eqref{eq:kparaCorrLen}].

For the particular computational set-up in this paper, one sees about a factor of 2 reduction in $N_\textrm{eff}$ for the highest $k_\perp$.  Going to even higher $k_\perp$ causes even greater reductions in sensitivity (compared to simple expectations).  The correlations discussed here are particularly important for experiments proposing to make measurements deep within the wedge.  As discussed above, the extent of the wedge in $k_\parallel$ provides the characteristic error correlation length between different $k_\parallel$ cells.  It therefore follows that errors are highly correlated whenever one chooses to work within the wedge.  Though such measurements may be well-motivated by the fact that the cosmological-signal-to-thermal-noise ratio is largest at low $k$, previous studies that established this have typically assumed the correlation length given by  Eq. \eqref{eq:kparaCorrLen}.  The resulting signal-to-noise ratios may therefore have been over-estimated.

We emphasize that the error correlations seen in this section exist whether or not one's power spectrum estimation pipeline includes a direct subtraction of modeled foregrounds from the input data.  To see this, suppose a direct foreground subtraction scheme reduces the foreground covariance by some constant multiplicative constant, so that $\mathbf{C}_\textrm{fg} \rightarrow \varepsilon \mathbf{C}_\textrm{fg} $ where $0 < \varepsilon \le 1$.  If one assumes that thermal noise is negligible compared to foreground residuals at low $k$ after a long time-integration, the result of our reduced $\mathbf{C}_\textrm{fg}$ will be a corresponding decrease in the amplitude of the final bias and the error bars.  However, the plots of error correlation in Figure \ref{fig:basicEstCovar} will remain unchanged, since the correlation is insensitive to an overall scaling.  The number of independent modes will therefore still decrease in the manner discussed above, reducing sensitivity.  Since this loss of sensitivity is the most pronounced at high $k_\perp$, it is particularly important to take into account for arrays that make use of long baselines, such as LOFAR or GMRT.

\section{Unsmooth foregrounds?}
\label{sec:Unsmooth}
In many prior works on $21\,\textrm{cm}$ cosmology, the assumption of spectrally smooth foregrounds is considered crucial to one's ability to perform foreground subtraction.  At various points in this paper, we too have assumed that foregrounds are smooth, and incorporating this assumption into our general framework gave rise to various predictions, such as the existence of the EoR window.  However, since the power spectrum estimation framework itself does not require smooth foregrounds, a number of our key results would survive a (hypothetical) discovery of unsmooth foreground sources.  We now briefly discuss how the problem of foreground mitigation would change if such an unfortunate discovery were to be made.

Because the window functions encode only the mapping of the true power spectrum to the estimated power spectrum and do not depend on the actual power spectrum, they do not rely on the assumption of smoothness.  Irrespective of whether the foregrounds are smooth or not, the window functions  accurately describe how foreground power is smeared out on the $k_\perp k_\parallel$ plane by our instrument.  If the foregrounds are smooth, their influence is limited to the wedge.  This makes foreground mitigation easy, as their avoidance requires no more than a simple cut on the Fourier plane.  If the foregrounds are not as smooth as expected, the EoR window will be smaller, but its exact size can still be predicted by convolving the (now unsmooth) model of our foregrounds with the same window functions as before.  Forthcoming data from various experiments at higher sensitivity will allow further foreground modeling, and---with the help of the window functions---an accurate determination of the extent of the foreground wedge.  Encouragingly, recent theoretical calculations have shown that in physically-motivated models of synchrotron emission, foreground spectra tend to be smooth even under the most pessimistic of assumptions \cite{Bernardi2014}.

For the sake of argument, however, let us consider a worst-case scenario where foregrounds are discovered to be sufficiently unsmooth for the EoR window to be drastically reduced in size.  In such a scenario, a number of strategies can be employed for foreground subtraction.  First, foregrounds can be modeled and subtracted to the best of one's ability in the visibility data.  Following that, a more sophisticated estimator (one that downweights the data not by $\mathbf{N}$ but by the total covariance $\mathbf{C}$ to account for uncertainty in the foregrounds) can be used.  Finally, the foreground bias can be subtracted from the power spectrum using Eq. \eqref{eq:WindToBias}, and window function decorrelation techniques can also be used in an attempt to increase the size of the EoR window.  We explore a number of these techniques in Paper II \cite{Liu2014b}.

\section{Conclusions}
\label{sec:Conclusions}
In any measurement of the redshifted $21\,\textrm{cm}$ power spectrum, foreground contamination is a serious concern.  Fortunately, observations and various theoretical studies have shown that despite complications arising from the inherently chromatic nature of an interferometric measurement, smooth spectrum foregrounds occupy a characteristic wedge region in cylindrical $k_\perp k_\parallel$ Fourier space.  The complement of this region is expected to be relatively foreground-free, forming an EoR window where measurements might be made.

While there exists an extensive literature on the topic, previous studies have typically focused on how the foreground wedge manifests itself in the mean power spectrum signal.  However, the same physical effects that cause the wedge in the power spectrum also affect the associated error statistics, such as the error covariance and the window functions.  An examination of some of these statistics was performed in Ref. \cite{Trott2012} using Monte Carlo methods.  In this paper, we have provided a complementary treatment by deriving a rigorous, fully-covariant mathematical description of the foreground wedge and the EoR window.  While our methods require the numerical \emph{evaluation} of some matrix expressions, they differ from previous work in that they do not require numerical \emph{simulations} of interferometric measurements, since the underlying framework is largely analytic.  This makes it possible to compute error statistics with very high dynamic range, which is crucial since the foregrounds are expected to dwarf both the instrumental noise and cosmological signal.

Our formalism takes advantage of the delay spectrum techniques introduced in Ref. \cite{Parsons2012b} to achieve computational savings, and in fact it is the use of the delay basis that makes our covariant, high dynamic range calculations numerically feasible.  However, we re-emphasize that this is merely a choice of basis, and that our results are independent of this choice.  This was shown explicitly in Section \ref{sec:basicEstAnalytic}, when we developed a description of the foreground wedge in terms of window functions.  Our description decouples the causes of the wedge---which depend only on the chromatic nature of the instrument and the specific form of our power spectrum estimator---from the detailed nature of the foreground emission.  Independent of foreground properties, window functions that are centered at high $k_\perp$ will typically develop long tails towards low $k_\parallel$.  The wedge then results from the additional assumption that foregrounds are spectrally smooth, so that strong signals from low $k_\parallel$ are transferred to higher $k_\parallel$ by the long tails.  Once the window functions have been computed, however, our formalism allows such assumptions to be relaxed.

With a fully covariant framework, we are able to track all error correlations in our numerical computations.  We find that measurements made at high $k_\perp$ have highly correlated errors, effectively reducing the number of independent measurements that can be made in that part of Fourier space.  This is particularly important for sensitivity forecasts that rely heavily on measurements made within the wedge, since the wedge's extent in Fourier space is roughly on the same scale as that of the error correlations.  Previous studies have typically neglected error correlations, assuming that errors are independent as long as the spatial Fourier cells are of the same size as an antenna's $uv$ footprint, and the spectral Fourier cells are on the order of $1/ B_\textrm{band}$.  Our work suggests that this is likely to be too optimistic an assumption.  At the highest $k_\perp$ considered in our numerical computations ($k_\perp = 0.13\,h$Mpc$^{-1}$), for example, error correlations reduce the number of independent modes by approximately a factor of 2.  This effect will be even more pronounced at even higher $k_\perp$, which are probed by experiments with extremely long baselines.  Since the chromatic effects that caused the wedge are closely related to those that cause error correlations, it will be crucial in future research to address the question of exactly how far the wedge can be pushed back (or equivalently, how much one can expand the EoR window).  In Paper II, we use the formalism of this paper to explore statistical methods for enlarging the EoR window \cite{Liu2014b}.

In this paper, our goal was to provide a rigorous treatment of the wedge.  Previous treatments have typically made different simplifying assumptions.  These include neglecting partially redundant baselines, approximating delay modes as $\eta$ modes, making assumptions about baseline length, assuming top-hat primary beams, neglecting binning artifacts, or assuming that errors are uncorrelated on the Fourier plane.  Our framework discards all of these approximations simultaneously, and it is gratifying to see that the basic picture of the EoR window as a naturally foreground-free region of Fourier space remains unchanged.  This bodes well for foreground avoidance efforts that aim to detect the EoR by working outside the wedge, making it possible for $21\,\textrm{cm}$ cosmology to open a new window into the high redshift universe using only existing data analysis techniques, with even more transformative results possible with further advances that expand the EoR window.

\section*{Acknowledgments}
It is a pleasure to acknowledge useful conversations with James Aguirre, Zaki Ali, Adam Beardsley, Chris Carilli, Josh Dillon, Aaron Ewall-Wice, Danny Jacobs, Matt McQuinn, Miguel Morales, Abraham Neben, Jonnie Pober, Jonathan Pritchard, Christian Reichardt, Katelin Schutz, Richard Shaw, Max Tegmark, and Chris Williams.  AL would like to additionally thank Peter Nugent and Casey Stark for their excellent course on high performance computing, as well as Allan Adams for his constant reminders that a simple trick of non-dimensionalization will often allow physical insight to be extracted from seemingly intractable integrals.  The authors would also like to thank the anonymous referee of this paper, whose helpful comments resulted in better (and more complete) explanations.  This research used resources of the National Energy Research Scientific Computing Center, which is supported by the Office of Science of the U.S. Department of Energy under Contract No. DE-AC02-05CH11231.  AL and AP were supported by NSF grants AST-0804508, AST-1129258, AST-1125558, and a grant from the Mt. Cuba Astronomical Association.  The Centre for All-sky Astrophysics is an Australian Research Council Centre of Excellence, funded by grant CE110001020. The International Centre for Radio Astronomy Research is a Joint Venture between Curtin University and the University of Western Australia, funded by the State Government of Western Australia and the Joint Venture partners.

\appendix
\section{Fourier conventions}
\label{appendix:Fourier}

In this Appendix, we define our Fourier conventions.  In sections where we establish formalism, we typically use a Fourier convention with factors of $2\pi$ in the exponent, so that the sky $I(\boldsymbol \theta, \nu)$ and its Fourier transform $\widetilde{I}(\mathbf{u}, \eta)$ are related by
\begin{equation}
\label{eq:Fourier1a}
\widetilde{I}(\mathbf{u}, \eta) = \int_{-\infty}^\infty I(\boldsymbol \theta, \nu) e^{-i 2 \pi (\mathbf{u} \cdot \boldsymbol \theta + \eta \nu)} d^2\boldsymbol \theta \,d\nu
\end{equation}
and
\begin{equation}
I(\boldsymbol \theta, \nu) = \int_{-\infty}^\infty  \widetilde{I}(\mathbf{u}, \eta) e^{i 2 \pi (\mathbf{u} \cdot \boldsymbol \theta + \eta \nu)} d^2\mathbf{u} \,d\eta,
\end{equation}
where $\mathbf{u}$ is the Fourier dual to $\boldsymbol \theta$ and $\eta$ is the Fourier dual to $\nu$.  Correspondingly, the power spectrum $P(\mathbf{u}, \eta)$ is defined as
\begin{equation}
\label{eq:PowerSpectrumAppendix}
\langle \widetilde{I}(\mathbf{u}, \eta) \widetilde{I}^*(\mathbf{u}^\prime, \eta^\prime) \rangle \equiv P(\mathbf{u}, \eta) \delta(\mathbf{u} - \mathbf{u}^\prime) \delta(\eta - \eta^\prime).
\end{equation}
Adopting this Fourier convention is convenient for developing a mathematical description of the foreground wedge because the Fourier transform closely mimics the definition of a visibility [Eq. \eqref{eq:basicVis}].  

Theoretical studies, however, typically use different coordinates and a different Fourier convention.  In cosmological coordinates, the Fourier transform is given by
\begin{equation}
\label{eq:Fourier1b}
\widetilde{T}(\mathbf{k}_\perp, k_\parallel) = \int_{-\infty}^\infty T(\mathbf{r}_\perp, r_\parallel) e^{-i (\mathbf{k}_\perp \cdot \mathbf{r}_\perp + k_\parallel r_\parallel)} d^2 \mathbf{r}_\perp dr_\parallel,
\end{equation}
and the inverse transform is given by
\begin{equation}
T(\mathbf{r}_\perp, r_\parallel) = \int_{-\infty}^\infty \widetilde{T}(\mathbf{k}_\perp, k_\parallel) e^{i (\mathbf{k}_\perp \cdot \mathbf{r}_\perp + k_\parallel r_\parallel)} \frac{d^2 \mathbf{k}_\perp dk_\parallel}{(2\pi)^3},
\end{equation}
where $T(\mathbf{r}_\perp, r_\parallel)$ is the sky temperature at comoving position $\mathbf{r}_\perp$ perpendicular to the line-of-sight and $r_\parallel$ parallel to the line-of-sight.  The power spectrum $\overline{P}$ is defined as 
\begin{equation}
\langle \widetilde{T}(\mathbf{k}_\perp, k_\parallel) \widetilde{T}^*(\mathbf{k}^\prime_\perp, k_\parallel^\prime) \rangle = (2\pi)^3 \overline{P}(\mathbf{k}_\perp, k_\parallel) \delta(\mathbf{k} - \mathbf{k}^\prime),
\end{equation}
where $\mathbf{k} \equiv (\mathbf{k}_\perp, k_\parallel)$.

Because the angular positions and frequencies can be mapped to transverse and line-of-sight comoving distances, respectively, the two Fourier conventions can be related to one another.  Defining $\mathbf{r}_\perp$ as the comoving distance perpendicular to the line-of-sight, we have
\begin{equation}
\mathbf{r}_\perp = D_c \boldsymbol \theta,
\end{equation}
with
\begin{equation}
D_c \equiv \frac{c}{H_0} \int_0^z \frac{dz^\prime}{E(z^\prime)}; \,\, E(z) \equiv \sqrt{\Omega_m (1+z)^3 + \Omega_\Lambda}.
\end{equation}
where $c$ is the speed of light, $z$ is the redshift of observation, $H_0$ is the Hubble parameter, $\Omega_m$ is the normalized matter density, and $\Omega_\Lambda$ is the normalized dark energy density.  The line-of-sight direction is more subtle.  When forming a power spectrum, one typically includes only data from a relatively narrow range in redshift.  Otherwise, cosmological evolution invalidates the assumption of a translation-invariant temperature field, which is needed in the definition of a power spectrum.  What matters, then, is not the mapping between frequency and the total comoving line-of-sight distance, but rather, the \emph{local} relation between differences in frequency $\Delta \nu$ and differences in distance $\Delta r_\parallel$ at the redshift of observation:
\begin{equation}
\Delta r_\parallel = \frac{c}{H_0 \nu_{21}} \frac{(1+z)^2}{E(z)} \Delta \nu,
\end{equation}
where $\nu_{21} \equiv 1420 \,\textrm{MHz}$ is the rest frequency of the $21\,\textrm{cm}$ line.  Having established this mapping we may (in what is perhaps an abuse of notation) recenter our coordinates so that $\Delta \nu \rightarrow \nu$ and $\Delta r_\parallel \rightarrow r_\parallel$.  Such a re-centering introduces a constant phase shift in our Fourier transforms, which has no bearing on quadratic statistics such as the power spectrum.

Making the identification between $T(\mathbf{r}_\perp, r_\parallel)$ and $I(\boldsymbol \theta, \nu)$ using our mappings, we may compare  Eqs. \eqref{eq:Fourier1a} and \eqref{eq:Fourier1b} to conclude that
\begin{equation}
\mathbf{k}_\perp = \frac{2\pi \mathbf{u}}{D_c}; \quad \mathbf{k}_\parallel =  \frac{2\pi \nu_{21} H_0 E(z)}{c(1+z)^2} \eta.
\end{equation}
This allows the power spectra defined under the different Fourier conventions to be related to one another:
\begin{equation}
\overline{P}(\mathbf{k}_\perp, k_\parallel) = \frac{c(1+z)^2 D_c^2}{\nu_{21} H_0E(z)} P(\mathbf{u},\eta).
\end{equation}
In this paper, we will plot all numerical results in terms of cosmological Fourier coordinates $\mathbf{k}_\perp$ and $k_\parallel$, even though we perform all our computations in terms of $\mathbf{u}$ and $\eta$.  We use WMAP9 cosmological parameters: $\Omega_m = 0.28$, $\Omega_\Lambda = 0.72$, and $H_0 = 69.7\,\frac{\textrm{km} /\textrm{s}}{\textrm{Mpc}}$ \cite{HinshawEtAl2013}.

\section{Equivalence of gridded and visibility-based approaches to basic power spectrum estimation}
\label{appendix:GridVsVisEquiv}

In this Appendix, we prove that estimating the power spectrum directly from visibilities using the basic estimator of Section \ref{sec:basicEstAnalytic} is equivalent to first gridding the visibilities in $uv\eta$ space and estimating the power spectrum by squaring and binning the results.

In Section \ref{sec:basicEstAnalytic} we considered a quadratic power spectrum estimator of the form
\begin{equation}
\widehat{p}_\alpha \propto \mathbf{x}^\dagger \mathbf{N}^{-1} \mathbf{C}_{,\alpha} \mathbf{N}^{-1} \mathbf{x},
\end{equation}
where $\mathbf{N}$ is the instrumental noise covariance matrix, $\mathbf{x}$ is the data vector (containing all spectral information from all baselines), and $\mathbf{C}_{,\alpha} \equiv \partial \mathbf{C} / \partial p_\alpha$ is the response of the covariance matrix to the $\alpha$th bandpower.  Working in the limit of continuous bandpowers and differentiating  Eq. \eqref{eq:genCovar} with respect to $P(u_\alpha, \eta_\alpha)$ gives
\begin{equation}
(\mathbf{C}_{,\alpha})_{ij} \equiv \frac{\partial \mathbf{C}_{ij}}{\partial P(u_\alpha, \eta_\alpha)} = h(u_\alpha, \eta_\alpha; b_i, \tau_i) h^* (u_\alpha, \eta_\alpha; b_j, \tau_j).
\end{equation}
This allows us to simplify our expression for the power spectrum:
\begin{equation}
\widehat{p}_\alpha \propto | \mathbf{h}^{\alpha \dagger} \mathbf{N}^{-1} \mathbf{x} |^2,
\end{equation}
where $\mathbf{h}^\alpha_i \equiv h(u_\alpha, \eta_\alpha; b_i, \tau_i )$.  If we imagine writing each $\mathbf{h}^\alpha$ vector as a column of a larger $\mathbf{H}$ matrix, the result can be compactly rewritten as
\begin{equation}
\label{eq:SqEst1}
\widehat{\mathbf{p}}_\alpha \propto \big{|} (\mathbf{H}^{\dagger} \mathbf{N}^{-1} \mathbf{x})_\alpha \big{|}^2,
\end{equation}
where we have similarly grouped the bandpowers into a vector $\widehat{\mathbf{p}}$.

We now show that essentially the same estimator results if one first uses the visibilities to form a (Fourier space) map of the sky, which is then squared to form the power spectrum.  Comparing  Eqs. \eqref{eq:genVtilde} and \eqref{eq:hDef}, we see that our measurement equation can be written as
\begin{equation}
\mathbf{x} = \mathbf{H} \mathbf{s} + \mathbf{n},
\end{equation}
where $\mathbf{x}$ is our data vector like before, $\mathbf{n}$ is the instrumental noise contribution to the data (with covariance $\langle \mathbf{n} \mathbf{n}^\dagger \rangle$ given by the instrumental noise covariance $\mathbf{N}$ discussed above), and $\mathbf{s}$ is a discretized ``map" of the Fourier sky, with elements given by $\widetilde{I}(\mathbf{u}, \eta)$ evaluated at predefined grid points.

In the generalized mapmaking problem, one seeks to use the data $\mathbf{x}$ to form an estimator $\widehat{\mathbf{s}}$ of the true sky $\mathbf{s}$.  This is accomplished in a lossless manner \cite{Tegmark1997a} by the estimator
\begin{equation}
\widehat{\mathbf{s}} = \mathbf{R} \mathbf{H}^{\dagger} \mathbf{N}^{-1} \mathbf{x},
\end{equation}
where $\mathbf{R}$ is an invertible matrix.  As discussed in Ref. \cite{Morales2008} the combination $\mathbf{H}^{\dagger} \mathbf{N}^{-1} \mathbf{x}$ constitutes a dirty map, and the role of $\mathbf{R}$ is to normalize and/or deconvolve this map.  If $\mathbf{R}$ is taken to be diagonal, then its role is merely one of normalization.  More complicated forms for $\mathbf{R}$ mix different pixels of the dirty map, in principle allowing the dirty map to be deconvolved.  For example, the choice $\mathbf{R} \equiv \left[ \mathbf{H}^\dagger \mathbf{N}^{-1} \mathbf{H} \right]^{-1}$ deconvolves the instrumental beam perfectly, giving an estimator with the property $\langle \hat{\mathbf{s}}  \rangle = \mathbf{s}$ (so that each pixel in the estimated map is on average probing only the corresponding pixel in the true map, rather than a linear combination of pixels in the pattern of a remaining point-spread function).

Suppose one forgoes deconvolution by picking $\mathbf{R}_{ij} \equiv r_i \delta_{ij}$, and then proceeds to form an estimate of the power spectrum by squaring the complex magnitude of the resulting map estimator $\widehat{\mathbf{s}}$.  The result is
\begin{equation}
\label{eq:SqEst2}
| \widehat{\mathbf{s}}_i |^2 = r_i^2 \, \big{|} \!\left( \mathbf{H}^{\dagger} \mathbf{N}^{-1} \mathbf{x} \right)_i \! \big{|}^2 \propto \widehat{\mathbf{p}}_i,
\end{equation}
thereby proving that the basic power spectrum estimator that we examined in Section \ref{sec:basicEstAnalytic} is equivalent to one where one forms a $uv\eta$ space dirty map from visibilities, and then squares the result.

We stress that the proof that we have just presented is basis-independent, in the sense that our data vector need not be indexed by baseline and delay.  For example, one may choose to deal with frequency spectra rather than delay spectra, in which case the data vector would be indexed by baseline and frequency channel.  The resulting $\mathbf{h}^\alpha$ vectors would no longer be given by  Eq. \eqref{eq:hDef}, but the proof shown here would be unchanged.

Crucially, the proof shown here assumed an infinitely-finely discretized Fourier space.  In practice, this will only be a good approximation on small scales (large Fourier wavenumbers), where the difference in wavenumber $\Delta k$ between neighboring discretized bins is small compared to the magnitudes of the wavenumbers $k$ themselves.  In Paper II we will formalize this assumption by importing the Feldman-Kaiser-Peacock approximation that is commonly used in galaxy surveys.

Importantly, we emphasize that while squaring a normalized $uv\eta$ dirty map is a perfectly reasonable way to estimate a power spectrum, it is by no means optimal.  Indeed,  Eqs. \eqref{eq:SqEst1} and \eqref{eq:SqEst2} are provably non-optimal, and better estimators are explored in Paper II.  Instead, the estimators considered here (and therefore in Section \ref{sec:basicEstAnalytic}) are intended to be representative of simple, ``first pass" methods \cite{Bernardi2013,Thyagarajan2013,Hazelton2013}, and their statistical properties provide basic pictures of the challenges that one faces.

\section{Technical computational details}
\label{appendix:Computational}

In this Appendix, we provide further details pertaining to the numerical computations described in Section \ref{sec:basicEstNumerical}.

\subsection{Binning}
In the quadratic estimator formalism used in this paper, there are two sets of discretizations: a discretization of the input data and a discretization of the output power spectra.

For all the computations performed here, we have chosen to express the input data in a basis parameterized by baselines and delay.  In other words, each component of the input data vector $\mathbf{x}$ corresponds to a different baseline and delay pair.  For computational tractability, we bin baselines together into $50$ linear bins of width $5\,\textrm{m}$, with the first bin centered about $10\,\textrm{m}$ and the last bin centered about $255\,\textrm{m}$.  (There are in fact some slightly longer baselines in our array.  However, we discard them in our analysis for reasons of numerical stability).  The delay bins are again linear, and range from $-200\,\mu\textrm{s}$ to $198.75\,\mu\textrm{s}$ in 320 bins.  This gives delay increments of $0.125\,\mu\textrm{s}$, equal to the natural bin size of $1/B_\textrm{band}$.  Note that this equality is by no means a requirement.  If computational resources are not a concern, it may be preferable to slightly over-resolve.  This is perfectly legitimate despite the fact that the bins will no longer be independent, since the inclusion of the channel profile $\gamma$ and our tracking of full covariance information allows the non-independence to be self-consistently captured.

Given that a baseline of length $b$ roughly probes modes with $u \sim b \nu_0 / c$, and that a delay bin $\tau$ roughly probes $\eta \sim \tau$, a logical way to discretize the output $u\eta$ space would be to use linear bins that matched the input bins (but with $u$ bins scaled by an appropriate factor of $\nu_0 / c$).  Such a scheme would be the most appropriate for matching the specifications of the instrument.  However, the cosmological power spectrum is expected to evolve on logarithmic $k$ scales.  Thus, a linear binning is computationally wasteful at high $k$, where a large number of bins are used to resolve a power spectrum that does not evolve very much.  On the other hand, a logarithmic binning scheme that is appropriate at high $k$ will tend to be computationally wasteful at low $k$, where one would be over-resolving the instrumental response.  As a compromise, we use a hybrid binning scheme that is roughly linear at low $k$ and roughly logarithmic at high $k$.  In this scheme, the $(n+1)$th boundary of the $u$ bins $u_{n+1}$ is given by
\begin{equation}
u_{n+1} = 1.036 u_n + 2.5.
\end{equation}
Similarly, the $(n+1)$th boundary of the $\eta$ bins $\eta_{n+1}$ is given by
\begin{equation}
\eta_{n+1} = 1.095 \eta_n + 0.125\,\mu\textrm{s}.
\end{equation}
At low $u$ and low $\eta$ the additive terms dominate, yielding bin boundaries that are spaced in an approximately linear fashion well-suited to the instrumental specifications.  At high $u$ and $\eta$ the multiplicative terms dominate, giving logarithmic bins that are a good fit for theoretical expectations.  For both $u$ and $\eta$, we use $30$ bins, giving a total of $900$ $u\eta$ bandpowers.  The bottom edge of the lowest $u$ bin is at $u=3$, while the bottom edge of the lowest $\eta$ bin is at $\eta=0.12\,\mu\textrm{s}$.

\subsection{Sparseness and computational shortcuts}

The methods and computations presented in this paper are basis-independent.  By this, we mean that while our final goal is to estimate a power spectrum (and its associated error statistics) on the $k_\perp k_\parallel$ plane, our input data may be expressed in any basis that we find convenient.  We now elaborate on our reasons for working in a baseline/delay basis.

As an example, consider the evaluation of $\mathbf{C}$.  With the Gaussian beams and tapering functions used in Section \ref{sec:basicEstNumerical}, the covariance $\mathbf{C}$ is given by  Eq. \eqref{eq:GaussianCovar}.  For parts of the matrix corresponding to short baselines, one can see by inspection that the matrix will be diagonal-dominant, with a large number of off-diagonal elements that are close to zero.  The $\mathbf{C}_{,\alpha}$ matrices are even more sparse, since many of the diagonal elements (those that do not satisfy $u \approx \nu_0 b_i / c$ or $\eta \approx \tau_i$) will also be zero.  In our computations, we skip the evaluation of matrix elements that are expected to be small.  This represents significant savings in computation time, given that with our binning scheme each matrix measures $16,000 \times 16,000$, and each element requires numerically integrating a two-dimensional integral given by  Eq. \eqref{eq:GaussianCovar}.  Moreover, with $900$ bandpowers, this process must be repeated $900$ times for each of the $\mathbf{C}_{,\alpha}$ matrices.

In our implementation, we set off-diagonal matrix elements of $\mathbf{C}$ to zero if the integrand is suppressed by at least $10^{-12}$ relative to the relevant diagonal elements everywhere over the integration volume.  For the $\mathbf{C}_{,\alpha}$ matrices, we apply the additional constraint that a diagonal element is to be skipped if the integrand is attenuated by $10^{-12}$ or more compared to a different diagonal element that satisfies $u_\alpha \approx \nu_0 b / c$ and $\eta_\alpha \approx \tau$, where $u_\alpha$ and $\eta_\alpha$ are the $u\eta$ values corresponding to the $\alpha$th band.

The sparseness that we have described here is a direct result of our using a baseline/delay basis.  In contrast, parameterizing the spectral information in a frequency basis results in substantially denser matrices, since the data are highly correlated between frequencies \cite{Trott2012}.  The delay transform roughly isolates spectral information by $\eta$ mode.  This isolation is imperfect in the long baseline limit, as we saw in  Eq. \eqref{eq:longBlVtilde}.  The matrices are therefore still dense for elements corresponding to long baselines, but the sparseness that is available with short baselines provides enough savings to enable the full propagation of covariant information without resorting to Monte Carlo methods, which can sometimes be slow to converge to the dynamic range displayed in Section \ref{sec:basicEstNumerical}.  Finally, we note that in an application of our methods to real measurements, the delay transform operates individually on each baseline \cite{Parsons2012b}, and therefore can be applied with negligible computational cost to the input data.

\bibliography{wedgeFormalismPartA}

\end{document}